\documentclass[12pt]{article}
\usepackage{astrobib,epsf,aaspp4,psfig,lscape}
\def\lap{\lower.5ex\hbox{$\; \buildrel < \over \sim \;$}}
\def\gap{\lower.5ex\hbox{$\; \buildrel > \over \sim \;$}}
\begin{document}
\title{Clues about the Star Formation History of the M31 Disk from WFPC2 Photometry\altaffilmark{1}}
\author{ Benjamin F. Williams}
\affil{University of Washington}
\affil{Astronomy Dept. Box 351580, Seattle, WA  98195-1580}
\affil{ben@astro.washington.edu}

\altaffiltext{1}{Based on observations with the NASA/ESA Hubble Space
Telescope obtained at the Space Telescope Science Institute, which is
operated by the Association of Universities for Research in Astronomy,
Inc., under NASA contract NAS5-26555.}

\begin{abstract}
Over the past several years, the Hubble Space Telescope (HST) has
acquired many broad-band images of various regions in the M31 disk.  I
have obtained 27 such fields from the HST data archive in order to
produce color-magnitude diagrams (CMDs) of the stellar populations
contained within these areas of the disk.  I have attempted to
reproduce these CMDs using theoretical stellar evolution models in
conjunction with statistical tools for determining the star formation
history which best fits the observations.  The wide range of
extinction values within any given field makes the data difficult to
accurately reproduce; nevertheless, I have managed to find star
formation histories which roughly reproduce the data.  These
statistically determined star formation histories reveal that, like
the disk of the Galaxy, the disk of M31 contains very few old
metal-poor stars.  The histories also suggest that the star formation
rate of the disk as a whole has been low over the past $\sim$1 Gyr.

\end{abstract}


\section{Introduction}

Galaxies are mainly classified by the spatial and color distribution
of their constituent stars; therefore, understanding the evolution of
galaxies requires the comprehension of their star formation histories
(SFHs).  The determination of such histories for several Local Group
dwarf galaxies has been of significant use for understanding the
circumstances under which these galaxies formed their stars (e.g.  And
I \cite{dacosta1996}, And II \cite{dacosta2000}, Carina
\cite{mighell1997}, Fornax \cite{buonanno1999}, the Large Magellanic
Cloud (\citeNP{holtzman1999} and \citeNP{olsen1999}), Leo I
\cite{Gallart1999}, Leo A \cite{tolstoy1998}, NGC 185
\cite{martinez1999}, Sculptor \cite{monkiewicz1999}, Sextans A
\cite{dohm1997}, Ursa Minor \cite{mighell1999}, WLM
\cite{dolphin2000b}, and many others).  One important example is the
relatively recent discovery that the cluster formation history and the
field star formation history of the Large Magellanic Cloud show
significant differences, forcing a re-evaluation as to how star
cluster formation and field star formation compliment each other in
the evolution of galaxies.  Measurements of the SFHs for Local Group
dwarf spheroidal galaxies have also proven interesting, showing that
these objects, once thought to contain a nearly coeval population,
actually have rather diverse evolutionary histories.  These observed
histories have been useful to compare with SFHs that come from models
of the Local Group to understand the effects of, for example, tidal
strirring, on galaxy evolution (e.g. \citeNP{mayer2001}).  Ideally one
would like to generate similar measurements of the star and cluster
formation histories of the large, more complex, spiral galaxy M31;
however, such a project is daunting due to M31's distance, size,
inclination, and internal reddening.

Due to the seeing limitations of ground based observations of M31,
deep photometry of its stellar populations was impossible before the
technology of space telescopes.  Integrated photometry provided some
interesting information.  For example, \citeN{dvc1958} found the
integrated color of the disk to be quite red (B-V $\sim$ 0.9),
suggesting that it was either heavily reddened or old or both.  The
limitation of achievable depth for stellar photometry meant that only
the most recent star formation was able to be understood.  Much
stellar photometry has been performed on the brightest stars (see
references in \citeNP{hodge1992}). More recently, the photometry of
brightest stars in many regions of the galaxy has been determined to
the best of the ability of ground based instruments by the studies of
\citeN{massey1998}, \citeN{hodge1988} and \citeN{magnier1997a}.  These
studies of the M31 disk population were limited in their conclusions
to the similarity of the most recent star formation history across the
disk from the upper main sequence (MS) luminosity function
\cite{hodge1988} and some evidence for spiral density wave propagation
across the southern half of the M31 disk from Cepheid age
determinations \cite{magnier1997a}.  \citeN{williams2001a}, using HST,
found that the ages of some of the largest clusters in the southern
region of the M31 disk were consistent with the ages of the field
Cepheids, providing further clues into the evolutionary history of the
M31 disk.

There has been a strong effort to use the abilities of the Hubble
Space Telescope in order to probe the older stellar populations in M31
and the stellar populations of the star clusters in M31 in an attempt
to constrain M31's star and cluster formation histories.
\citeN{holland1996} found evidence for multiple populations in the M31
halo using deep V and I band WFPC2 photometry, showing that this kind
of work is possible with the resolution of HST. Futher progress has
been made very recently with work by \citeN{sarajedini2001} who found
strong evidence for a thick disk population in an HST field at a
smaller galactocentric distance than that studied by
\citeN{holland1996}.  While all of these studies were able to put
useful broad constraints on the general properties of the populations
in these fields and clusters, they did not provide a statistically
determined SFH for the fields.

In this paper, I make a step toward understanding the field star
formation history of M31 using the deepest photometry available from
the HST archive.  I perform photometry on 27 WFPC2 fields taken from
the HST archive: 13 fields in B and V and 14 fields in V and I.  I
apply a statistical analysis to the stellar photometry, comparing the
color-magnitude diagrams to CMDs generated from theoretical stellar
evolution models in order to find the best fitting SFH for the fields.
The results provide peeks at the SFH of the disk as a whole through
the limited field of view of HST.  Section 2 describes the method used
for finding and retrieving the photometry from the fields, including
artificial star tests.  Section 3 describes how I use this photometry
to determine the most likely SFHs of the fields.  Section 4 discusses
the many experiments which were done in order to find the SFH which
best reproduced the data.  Section 5 presents the results from the 27
fields including the similarities and differences of the SFHs in
different regions around M31.  Finally, section 6 gives some general
conclusions which may be drawn from the results.

\section{Data Acquisition and Reduction}

I searched the HST archive for all exposures longer than 60 seconds
which were taken through the broadband B (F439W), V (F555W), and I
(F814W) filters pointing within 1.5 degrees of the M31 nucleus.  Using
this narrow radius kept the data contained within the disk, avoiding
significant halo contamination.  Any fields in the bulge were later
excluded by eye.  I acquired and reduced 27 WFPC2 fields from the HST
archive, each observed through either the B and V broad-band filters
or the V and I broad-band filters.  These field were put together
using the data listed in Table 1 (electronic version only), which
gives the field name, proposal identification number, RA, DEC, date,
bandpass, and exposure time for each image taken from the HST
archive. The westernmost field in the sample was the only one taken
outside the 1.5 degree cutoff because it was located near the major
axis 2 degrees to the southwest of the center of the galaxy, allowing
a comparison of the outer and inner disk populations.  The positions
of these fields on the galaxy are shown in Figure 1.  This figure shows
an H$\alpha$ mosaic of the M31 disk \cite{winkler1995} with the
positions of 26 of the fields (the OUTER field is off the image to the
southwest) marked with squares showing the positions of the WF1, WF2,
and WF3 chips in the field.  In order to avoid special objects often
contained in the PC of each field, I excluded the PC data from the
analysis.  I also excluded the WF3 chip of the G272 field, since a
large portion of it was contaminated by a saturated bright star.  Each
field is labeled with the name it was given in Table 1 (electronic
version only). These names were taken from the image header field
TARGNAME when the images arrived from the data archive.  The random
distribution of most of the fields suggest that conclusions can be
drawn reliably and are not likely due to selection effects.

I determined the Johnson B, V and I magnitudes of the stars in these
frames using the automated photometry routine HSTPHOT
\cite{dolphin2000} which is optimized for detecting and measuring
magnitudes of stars on undersampled images like those taken with the
WFPC2 camera.  This photometry package disregards all saturated and
hot pixels using the data quality image provided by the data archive
and simple statistical tests.  Then it uses the stars in the frame to
make minor adjustments to a well-sampled, previously determined point
spread function (PSF).  This PSF is then fit to the brightness peaks
in the data.  Objects that fit the PSF are classified as stars and
their instrumental magnitude is measured.  These instrumental
magnitudes are finally converted to final ground based B, V, and I
magnitudes by applying aperture corrections determined for the PSF
using photometry from less crowded fields, applying a charge transfer
efficiency correction depending on chip position, and applying the
transformation equations and zero-points determined by
\citeN{holtzman1995}.  The CMDs obtained from the final photometry for
these fields are shown in the upper left panels of Figure 5 (a-z,
aa-ac; some parts are electronic only).

Once the photometry had been measured, artificial stars were added to
the image and the photometry routine was run on the image again in
order to determine the completeness and photometric errors as a
function of magnitude.  The results of the artificial star tests for
each field are shown Figure 2 (or Figures 2a-2c, electronic version
only); since the artificial star tests were run iteratively, there
were hundreds of thousands of artificial star results.  Each figure
shows a random sampling of 1 percent of these results in order to keep
the figure easy to interpret.  The M31 disk is a complex background
with which to work and therefore limits the accuracy of the photometry
by increasing the uncertainty of the local background level.  This
uncertainty is partially due to the Poisson noise of the higher
background levels, but it is also due to actual structure in the
background on spatial scales relevant to stellar photometry.  These
surface brightness fluctuations likely come from crowded, faint stars
in disk which cannot be resolved by HST; dust can also contribute to
these fluctuations.  The artificial star tests showed that despite
these problems, HST is capable of supplying photometry accurate to
$\sim\pm$0.2 magnitudes at $B=24, V=25$, and $I=24$.  I found
that, since the V-band images were usually deeper than the B-band, the
B-band artificial stars at the faint end in the B-band were only
picked up when they appeared on noise spikes.  This effect resulted in
the faint stars being measured to be bluer in B-V, causing the faint
blue tail seen in the V, B-V CMDs in Figure 5.

\section{Determining Star Formation Histories}

I determined the SFH and chemical evolution history which best
reproduced the CMD for each field using the statistical CMD matching
package MATCH \cite{dolphin1997}.  This recently developed software
uses a technique pioneered by earlier, non-computationally intensive
work on stellar populations in the Magellanic Clouds
(e.g. \citeN{gallart1996}) in which the CMD of the stars was broken
down into bins of color and magnitude.  These binned CMDs are also
known as Hess Diagrams.  Before powerful computing was possible, these
bins were chosen to isolate populations of differing ages and
metallicities, and based on the relative number of stars in each bin,
a rough measurement of the star formation and chemical enrichment
histories of the field could be made.  MATCH takes this concept to its
full potential by using the most recent computing power to create
high-resolution Hess Diagrams of rectangular bins of constant size
specified by the user.

Once the CMD of the data has been transformed into a Hess diagram, the
software uses the stellar evolution models of \citeN{girardi2000} to
create a unique Hess diagram for each of a range of stellar ages and
metallicities.  These model-based diagrams are produced by populating
the CMD along the theorectical isochrone of each metallicity and age,
assuming a 1 $M_{\odot}/yr$ star formation rate and a Salpeter initial
mass function (IMF) and taking into account the completeness and
photometric errors in each bin of the CMD as measured from the
artificial star tests on the real data.  The program then finds the
linear combination of model-based Hess diagrams which best reproduces
the Hess diagram of the observed stars for each of a range of
reddening and distance values.  Since each model-based diagram was
created assuming a 1 $M_{\odot}/yr$ star formation rate, the
coefficients of this linear combination provide the star formation
history in units of $M_{\odot}/yr$.  Finally, the statistically best
fits for each distance and reddening value are weighted by goodness of
fit to determine values and errors for the most likely distance and
extinction to the field as well as values and errors for the star
formation rates and metal abundance spreads during each time period
explored.  Finally, in order to check the viability of the result, the
package can be used to create a synthetic CMD from the stellar
evolution models by populating the theoretical isochrones using the
assumed IMF along with the derived best fit for the metallicities and
star formation rates for each time period, taking into account the best
fit distance and reddening values for the field.  The artificially
generated CMD can then be compared to the observed CMD in order to
verify that the statistically determined SFH creates a stellar
population similar to the observed population.

I have created artificial CMDs for each timestep individually for one
of the fields which was observed in B, V, and I (NGC224-FIELD2).  I
show these CMDs along with the SFHs which created them in Figures 3a
and 3b. These CMDs show how the different areas of the CMD are
populated by the different age groups.  Figure 3a shows the areas of
the V, B-V CMD which are filled in by the different timesteps.  It
shows that the V, B-V CMD is not reliable for determining the early
star formation rate of the field ($\gap$1 Gyr).  Due to the poor
sensitivity of WFPC2 in the blue, the red giant branch (RGB) is not
well sampled in this color so that the SFH of the old stars cannot be
reliably constrained.  On the other hand, the I, V-I CMDs are quite
sensitive to the older population.  These CMDs provide more reliable
constraints on the early star formation rate, and they also show a
recent SFH consistent with that determined from the B and V
photometry, as discussed in section 5.2.

\section{Discussion}

There was one serious problem with attempting to recreate the
photometry of these disk fields: the spread in extinction values
surely present due to the stars within a given field residing at
different positions in the disk.  Since some of the stars in these
fields were more heavily extincted than the others, the photometric
errors upon which the MATCH software was basing its statistics were
only part of the true scatter in the CMDs.  I ran many experiments
using the MATCH software in order to find the optimal parameter sets
for these fields.  The experiments were run using different parameter
combinations on the same dataset in order to find the parameters whose
resulting artificial CMD appeared to match the strongest features of
the observed CMD for that field.  In order to force the program not to
output non-physical chemical evolution and SFHs by fitting the stars
scattered about the strong features of the CMD, I was forced to place
limits on the parameter space by iteratively running the MATCH
software, limiting the parameter space as I went along.

First I noticed that exploring a large range of distances and
reddening values led to problems.  With such freedom of parameter
space it was not possible for the software to untangle the competing
effects of distance, reddening and metal abundance.  Differential
reddening could cause the software to find non-physical fits to the
scatter of stars to the red and blue of the RGB which are surely MS
and RGB stars with extinction well above the mean.  To stop the
software from trying to fit these stars, it was necessary to limit the
ranges of distances and reddening values used for the fits attempted
by the software.  This limitation made it impossible for those areas
of the CMD to be fit, forcing the software to find the best fit
possible without considering the bulk of the heavily reddened stars.
In order to perform this limitation of parameter space in a logical
fashion, I used knowledge of the reddening values for the fields in
the arms (all fields observed in B and G76) from studies of the star
clusters \cite{williams2001b}, knowledge of the foreground reddening
from Galactic reddening maps \cite{burstein1982} and the relatively
well constrained distance to M31 to set tight constraints on these
free parameters.  I allowed distance moduli from
24.35$\leq$m-M$\leq$24.55.  This small range allowed the software to
accomodate any slight systematic shifts in the positions of the
strongest CMD features, while forcing the distance to be consistent
with published measurements.  We were not interested in using these
data to measure the distance to M31.  Instead, we used the
well-constrained previous distance measures to M31 to limit the
parameter space searched by the software.  The distance moduli all
came out the same within the errors, which showed that our photometry
was very consistent and that the statistical fits were working.  I
allowed extinction values from 0.1$\leq$A$_V\leq$0.8 for the interarm
fields and from 0.1$\leq$A$_V\leq$1.2 for the fields within the arms.
The distance and reddening values determined for the fields are given
in Table 2.

More experiments also showed that it was necessary to restrict the
metallicity range the software could explore.  I found that most
results on a given set of stars contained the same basic star
formation history, but that the metallicity range which I allowed the
software to search had a significant impact on the resulting
artificial CMD.  For example, if the metallicity parameters were too
broad, the best fitting chemical evolution history would often be
discontinuous, containing large metallicity jumps in the abundance
history which could not be physical.  These jumps were likely
providing the best fit to the few stars scattered to the blue of the
RGB, but these seemingly impossible results always created patterns in
other parts of the artificial CMD which did not appear in the data,
such as gaps in the RGB.  Since the old stellar components in the
formation history were statistically best fit by a metal rich
population ([Fe/H] $\gap$ -0.5), I found better artificial CMDs were
created from fits which did not allow the metallicity to drop below
-1.0 dex, stopping the metallicity from jumping around too
drastically.  This limitation was imposed on the parameters only after
experiments allowing lower metallicities showed that the best-fit
abundances of the oldest stars were only a few tenths below solar.
The need to restrict the chemical evolution parameters to the
metal-rich end provides the first hint that the M31 disk, like that of
the Galaxy, contains very few old metal poor stars.

As a final experiment for discovering the accuracy of the SFH
measurements, I created a set of artificial star photometry with a
constant star formation rate and a linear increase in metallicity from
[Fe/H] = -2.0 at t = 15 Gyr to [Fe/H] = -0.1 at t = 0.1 Gyr.  I ran
this artificial photometry through the SFH software.  The results are
shown in Figure 4.  Panels (a) and (b) show the results on the
experimental field before the differential reddening was added.  In
this case, the analysis routine was sensitive to the old metal poor
population; however, it found the recently formed stellar population
to be [Fe/H] $\sim$ -0.5 instead of solar.  This result confirmed my
suspicion that I could not accurately determine metallicities from the
upper MS.  Panels (c) and (d) show the results when I altered the
artificial photometry by giving each star a unique reddening value,
$A_V\pm0.3$, away from the mean reddening for the field.  Fortunately,
the resulting SFH was not significantly altered from the input
history.  In both cases, the program found a nearly constant star
formation rate and the old metal poor population.  This result gave me
further confidence that the routine was finding useful SFHs, even
though the abundances were a more difficult problem.  In the case with
the simulated differential reddening, the program was even farther
from the correct metal abundances for the most recent star formation.

These experiments revealed that the abundances determined for the most
recent star formation ($\lap$100 Myr) were not robust since the color
of the upper MS is not sensitive to metallicity.  Also, the
region of the CMD used to determine star formation rate and metal
abundance for the most recent time periods was often sparsely
populated, meaning that the metal abundance determination for these
periods was based on small number statistics.  Therefore, in many
cases, I artificially increased the abundances of the most recently
formed stars in order to avoid non-sensical jumps in the abundance
history and in order to be consistent with spectra of M31 HII regions
and supernova remnants which show that recently formed stars in the
M31 disk are formed from gas with abundaces similar to that of the
Orion nebula \cite{blair1982}.  This artificial increase in metal
abundance for the most recent time periods had the only effect effect
on the resulting artificial CMDs of removing an unobserved clump of
stars from above the observed tip of the RGB in the V, B-V CMDs.  The
increase had no noticeable effect on the artificial I, V-I CMDs.

\section{Results}

Four-panel figures are shown in Figure 5(a-z, aa-ac; some figures are
only shown in the electronic version of the publication).  The upper
and lower left panels show the observed and artificially generated
CMDs.  The artificial CMD was generated using the SFH and chemical
enrichment history shown in the upper and lower right panels along
with a Salpeter IMF to populate the stellar evolution models of
\citeN{girardi2000}.  The star formation rates and metal abundances
are shown as a function of log time because the stellar evolution
models are done with logarithmic time resolution.  As we try to
untangle the SFH of a field, the time resolution we can hope to
achieve decreases as we go farther back in time.  In fact, the time
bins between 1 and 10 Gyr are of higher resolution than I can actually
obtain from these data, which is likely the reason for the near
constant star formation rates in the early time bins.  This constant
rate more likely corresponds to an average of the star formation rate
from 1-10 Gyr.  The star formation rates and abundances derived from
the B and V data for early time periods ($\gap$1 Gyr) are not reliable
because the older population is not well sampled by these colors.  The
artificial CMD shown for each field is the closest match I was able to
find by running many experiments on the data using the MATCH software.
Here I discuss the implications of characteristics of the SFHs as well
as some fields with unique characteristics.

\subsection{General Trends}

There are several traits common to the SFHs of most of the disk
fields.  These global traits reveal stellar populations that are
present throughout the disk as well as possible systematic problems
with the analysis.  The V and I data show the general activity of the
disk early in its formation, while the B and V data, which are
generally restricted to the active arms, show the general activity of
the presently most active portions of the disk.

The most obvious common trait to the fields observed in V and I,
sensitive to the early SFH, is their similar star formation rates and
abundances early in the formation of the disk.  The star formation
rates from 10-15 Gyr ago are consistently between 0.001 and 0.01
M$_{\odot}$/yr for all M31 disk fields studied in V and I.  In
general, this early rate has not been surpassed recently except in the
most active spiral arm fields.  The implication of such a widespread
high star formation rate early is that the disk was almost surely more
active in its early history than in more recent history.  Assuming all
of the pointings which were observed in V and I provide a typical
sample of the early disk population and that any other HST field in
the disk will show the same early star formation rate as those already
observed, I can make a rough determination of the early star formation
rate of the M31 disk as a whole.  If the main disk covers an area of
$\sim$3 degrees$^2$ and one WFPC2 field covers an area of $\sim$4.7
arcmin$^2$, then the total disk SFR would have been $\sim$2-20
M$_{\odot}$/yr.  This rate for the early disk is larger than the
current rate for the entire galaxy. The total star formation rate at
present could not be calculated from our data because the young stars
are not as uniformly distributed throughout the disk.  This
non-uniform distribution invalidates the assumption that the current
rates for all HST fields in the disk will be the same, and in fact,
they are very different for each field.  Fortunately, the current star
formation rate for the entire disk can be estimated from the far IR
luminosity of 1.5$\pm$0.15 $\times$10$^9 L_{\odot}$
\cite{devereux1994}.  Using the method of \citeN{inoue2000}, the
resulting current SFR is $\lap$1 M$_{\odot}$/yr, clearly less than
that measured for the early disk.

Another trait common to the fields outside of the spiral arms (all
fields observed in I except NGC224-FIELD1, NGC224-FIELD2, and G76) is
a drop in the star formation rate at $\sim$1 Gyr.  This general trend
is consistent with the previous calculation that the present total
star formation rate in the disk is lower than it was at early times.
These results all point to M31 having undergone an overall decrease in
activity over the past Gyr, perhaps revealing that M31 is in a
relatively quiescent phase of its evolution.  This decrease is not
seen in the fields observed in B and V, which are more sensitive
probes of the recent SFH; however, there is a tendency for observers
to use these colors to study the active spiral arms in order to study
massive star formation.  The two fields observed in all three bands
are also in currently active regions and do not show this decline.

The fields in the arms tend to be observed in B and V, and the recent
SFH of these fields suggest that no significant overall change in
activity has occurred in the arms in the past Gyr.  All the fields
within the NE spiral arm show a fairly constant rate of star formation
over the past few Myr.  This result is in good agreement with previous
work on the age distribution of the star clusters in these fields
which suggests that there are clusters of all ages less than a few
hundred Myr in these fields \cite{williams2001b}.  Since the time
resolution is limited and differential reddening tends to broaden the
MS, this constant rate does not exclude the possibility of significant
short-timescale activity changes ($\lap$50 Myr).

Some common characteristics of the data reproductions provide
diagnostics for problems with the analysis technique.  There is a
tendency for the artificial CMDs to have tighter features than those
seen in the data.  For example, the RGB in the G11 field (Figure 5t)
is much broader in the data than in the artificial CMD, and the MS in
the NGC224-VDB-OB042-POS02 (Figure 5w) field is also much broader than
in the artificial CMD.  This difference is most likely due to the
range of extinction values for the stars in these fields.  This
differential reddening causes broadening of the MS and RGB which
cannot be fit by the stellar evolution models.  I have taken great
care to systematically fit the main features of the CMD in order to
avoid falsely mistaking stars with different reddening values to be
different age populations.  The non-recovery of the full width of
these features shows that the stars that suffered from extinction much
different from the mean extinction were not included in the SFH.
There is also a tendency for the bluer RGB stars to be fit by a
younger, low metallicity giant branch.  This fit is likely incorrect
because the branch created extends to brighter magnitudes than are
present in the data and because the sharp, significant drop in
metallicity cannot be physical.  More likely, these bluer RGB stars
are of similar metallicity to the rest of the RGB but are experiencing
less than the mean extinction.  The errors associated with this
reddening issue could not be quantified.  These errors are not
included in the figures, and therefore the quoted errors are likely to
be somewhat under-estimated.

\subsection{Specific Fields}

Some of the fields in the sample had special characteristics.  Two
fields were observed through all three filters, allowing tests of the
consistency of the results from the two colors.  Two other fields
contained nearly 100\% overlap, allowing tests of the consistency of
the method applied to images of the same field taken at different
times where the stars lay in different positions on the chips.  One
field has been previously studied using more traditional methods,
allowing a comparison of results.  Finally, one field lies far out in
the disk, allowing a comparison with the inner disk populations.
These special fields were useful for reaching an understanding of the
effectiveness of the analysis.

Two of the fields (NCG224-FIELD1 and NCG224-FIELD2) were observed in
B, V and I, allowing a comparison of the results obtained from the
field from the different colors.  The history determinations for each
field in B and V were consistent with the determinations in V and I
for recent times.  Due to the poor sensitivity of WFPC2 in the B band,
the results for the early times from the B and V data are not
reliable.  The regions of the CMD which probe the early time periods
are too sparsely populated to make reliable star formation rate
determinations.  On the other hand, the V and I data were useful for
constraining both the early and recent SFH.  These fields were two of
the most poorly reproduced V, V-I CMDs of the sample.  One important
reason for this poor fit was the large number of observed bright stars
scattered outside of the typical features of the CMD.  The three V,
V-I fields located in the spiral arms all had a large number
($\gap$100) of these stars, high mean reddening, and poor CMD fits.
These stars are very young, and they are likely more strongly
differentially reddened.  These problems together make the fields much
more difficult to decipher because the stellar evolution models are
not well understood for these massive stars and differential reddening
further confuses the situation.  The distance and mean reddening
determinations were consistent in NCG224-FIELD2 but not in
NCG224-FIELD1. The mean reddening determined from the V and I data was
lower than that obtained from the B and V data in NCG224-FIELD1.  Most
likely, the younger MS stars in this field are slightly more reddened
on average than the older RGB stars.  The consistency of the results
in general for the two fields is reassuring that the errors calculated
for the formation rates are not largely under-estimated for the fields
observed in V and I, and are not largely under-estimated back to
$\sim$1 Gyr for fields observed in B and V.  The inconsistency between
the metallicity results of the most recent time bins of the same
fields in the two colors reinforces the suspicion that the errors for
the metallicities determined for the recent star formation are largely
under-estimated.  These metallicity measurements rely heavily on the
evolution models for massive stars, which are still debated and could
have problems (e.g. \citeNP{limongi2000}; \citeNP{heger2000a}, and
references therein).

The fields G287 and G11 are extremely close together.  There is no
noticeable difference between their positions in Figure 1, and their
pointings are only a few arcseconds different.  I reduced and
determined the SFH of the two fields independently.  The results are
shown in Figures 5s and 5t respectively.  The SFHs determined for the
two fields are perfectly consistent, confirming the accuracy of the
photometry as well as the stability of the resulting SFH.

The field G272 has recently been analyzed by \citeN{sarajedini2001}.
They determined the metallicity distribution function of this field
concluding that the field was generally metal rich [Fe/H] $\sim$ -0.1.
I, too, find that the field is dominated by a metal rich population of
similar abundance to what they found.  In addition to the agreement
with \citeN{sarajedini2001} about the dominance of a metal rich
population in the G272 field, I also independently discovered the lack
of a population with age $\lap$1 Gyr.  According to this analysis,
there was a sharp decrease in star formation beginning at $\sim$1 Gyr.
These traits of this field were suggested as evidence that it was
dominated by a thick disk population.  Since this intermediate age,
metal rich population is detected in all of the fields sensitive to
it, this population is likely present throughout the disk, as one
would expect if it represents a thick disk population.
\citeN{sarajedini2001} also found a metal poor tail down to [Fe/H]
$\sim$ -2.5.  The MATCH analysis was not sensitive to such a metal
poor tail.  This tail is more prominent in the M31 halo
\cite{holland1996}, suggesting that it represents the halo population,
which would not be prominent in any of these fields.  Another
difference between our results is in the population of the tip of the
RGB.  \citeN{sarajedini2001} found only a handful of RGB stars of $I
\lap$ 22 whereas I find 1529 stars with $I \leq$ 22.  While this
difference is quite noticeable in the bright end of the CMD, it did
not significantly alter the conclusions drawn from the data.  I
attempted to recreate their lack of bright stars by reducing the field
using a different software package, DAOPHOT II \cite{stetson}, but the
resulting CMD was not significantly different from that obtained from
HSTPHOT.  The reason for the discrepancy remains a mystery.

For comparison purposes, I analyzed the OUTER field, which lies much
farther out in the disk than the other fields in this study.  This
field is found to be metal poor compared to the disk fields closer in,
disregarding the most recent time bins which have very few stars upon
which to determine the metallicity and should not be trusted.  This
lower metallicity in the OUTER field is consistent with findings from
abundance studies of emission nebulae in the M31 disk which find
evidence for an abundance gradient within the disk
(e.g. \citeNP{walterbos1999}).  There is also a surprisingly evident
population of young stars at $\sim$100 Myr.  The age of this minor
burst in star formation in the outer disk corresponds remarkably well
with the ages of may of the Cepheids in the southern disk
\cite{magnier1997a} as well as the ages of the massive and compact
clusters in the southern disk \cite{williams2001a}.  The early star
formation rates are also slower than those of all fields located
closer to the galaxy center, revealing the generally less active
nature of the outer disk.  Surprisingly, the old, very metal-poor tail
that represents the halo population is not detected in this outer disk
field either.  This non-detection serves as a warning that the
abundance determinations for the early history may suffer from
systematic problems resulting from uncertainties in the stellar
evolution models.

\section{Conclusions}

I have performed photometry on the stars contained within the wide
field chips of 27 WFPC2 fields in the M31 disk.  These fields were all
observed through broadband filters, and they were obtained from the
HST data archive.  Using these data, I have put constraints on the
SFHs of these portions of the M31 disk.  Taken together, several
general conclusions can be drawn from these SFHs: some pertaining to
the limitations of this method, and some pertaining to the evolution
of the M31 disk.  The many experiments, the work with simulated data,
and the random distribution of most of the fields suggest that these
conclusions can be drawn reliably and are not likely due to systematic
errors or selection effects.

By comparing the results of the same field analyzed via two colors
independently, I found that the overall results from the method are
stable, but the early SFH is better determined from V and I data.  The
differences in the reddening values measured for the same field by the
two colors could be explained by the younger stars experiencing a
different mean reddening than the older population.  The differences
in metallicity measured for recent times by the two colors could be
attributable to problems with the models in predicting the rapid color
changes that occur during the evolution of massive stars.  The errors
of these measurements are based on photometric errors from artificial
star tests which do not account for the dispersion caused by
differential reddening or problems inherent to the stellar evolution
models.  These unpredictable sources of confusion inevitably lead to
under-estimated errors.  Consistent results for the SFH of the two
fields observed in two colors and from significantly overlapping
fields suggests that these results are able to be reproduced, that the
errors on the star formation rates are not largely under-estimated,
and that the errors quoted for the photometry are accurate.

Overall, the SFHs determined for these 27 fields show some interesting
similarities and differences.  Comparison of these results with those
of previous studies of a common field reveals that most of these disk
fields sensitive to the old stellar populations contain a significant
intermediate-age, metal-rich population possibly associated with a
``thick disk'' in M31.  This population is joined by a significant
younger population in fields within spiral arms (G76, NGC224-FIELD1,
NGC224-FIELD2). This stellar population is not joined by a younger
population between or outside the spiral arms (G33, G87, K108, G119,
NGC224-DISK, NGC221-POS2, INNER, G11, G287, G272, G322).  All of the
fields sensitive to the old stellar populations (observed in V and I)
show a similar star formation rate and metallicity early in the
history of the disk (except the OUTER field, which lies much farther
away from the galaxy center than the rest of the sample).  The similar
rates suggest that the M31 disk as a whole was more actively forming
stars early in its evolution; the common high metallicities suggest
that, like the disk of the Galaxy, the M31 disk is deficient in old
metal poor stars.  At the same time, the differences amoung the recent
star formation rates with position within the disk suggest that over
the past Gyr some areas of the disk have become less active while
other areas have sustained a similar star formation rate to that of
the early history of the disk.  Since the old stars have certainly
been well mixed since their formation, there is no way to know if the
early star formation was more global or more intense.  These results
merely suggest that the total star formation rate for the disk was
significantly greater until about 1 Gyr ago.  One interpretation of
these results is a general evolution of the M31 disk consisting of
very active star formation ($\sim$2-20 M$_{\odot}$/yr) until about 1
Gyr ago, when the overall star formation rate declined with the
exception of some regions which retained or aquired sufficient gas for
continued or episodic star formation.

\section{Acknowledgments}

I thank Paul Hodge for his advice in dealing with the many problems
faced during this work and for his help in preparing the manuscript.
I thank Andrew Dolphin for helping with the MATCH and HSTPHOT
packages.  Support for this work was provided by NASA through grant
number GO-08296 from the Space Telescope Science Institute,
which is operated by the Association of Universities for Research in
Astronomy, Incorporated, under NASA contract NAS5-26555.


\begin{landscape}
\begin{deluxetable}{ccccccc}
\small
\tablewidth{8.0in}
\tablecaption{Data obtained from the HST data archive used for the cluster survey.}
\tableheadfrac{0.05}
\tablehead{
\colhead{\bf{Field}} &
\colhead{\bf{Prop. \#}} &
\colhead{\bf{Obs. date}} &
\colhead{\bf{RA (2000)}} &
\colhead{\bf{DEC (2000)}} &
\colhead{\bf{Filter}}  &
\colhead{\bf{Exp. (sec)}} 
}
\startdata
OUTER & 6859 & Aug 14 1997 &  0:36:59.20 & 39:52:21.3  &   F555W &  300\nl
OUTER & 6859 & Aug 14 1997 &  0:36:59.20 & 39:52:21.3  &   F555W & 1000\nl
OUTER & 6859 & Aug 14 1997 &  0:36:59.20 & 39:52:21.3  &   F555W & 1100\nl
OUTER & 6859 & Aug 14 1997 &  0:36:59.20 & 39:52:21.3  &   F814W &  200\nl
OUTER & 6859 & Aug 14 1997 &  0:36:59.20 & 39:52:21.3  &   F814W & 1000\nl
OUTER & 6859 & Aug 14 1997 &  0:36:59.20 & 39:52:21.3  &   F814W & 1300\nl
G33 & 6671 & Feb 26 1999 &  0:39:32.23 & 40:30:48.0  &   F555W & 1200\nl
G33 & 6671 & Feb 26 1999 &  0:39:32.23 & 40:30:48.0  &   F555W & 1400\nl
G33 & 6671 & Feb 26 1999 &  0:39:32.23 & 40:30:48.0  &   F555W & 1300\nl
G33 & 6671 & Feb 26 1999 &  0:39:32.23 & 40:30:48.0  &   F555W & 1400\nl
G33 & 6671 & Feb 26 1999 &  0:39:32.23 & 40:30:48.0  &   F814W & 1300\nl
G33 & 6671 & Feb 26 1999 &  0:39:32.23 & 40:30:48.0  &   F814W & 1400\nl
G33 & 6671 & Feb 26 1999 &  0:39:32.23 & 40:30:48.0  &   F814W & 1300\nl
G33 & 6671 & Feb 26 1999 &  0:39:32.23 & 40:30:48.0  &   F814W & 1400\nl
G38 & 8296 & Oct 15 1999 &  0:39:47.35 & 40:31:57.9  &   F555W &  600\nl
G38 & 8296 & Oct 15 1999 &  0:39:47.35 & 40:31:57.9  &   F555W &  600\nl
G38 & 8296 & Oct 15 1999 &  0:39:47.35 & 40:31:57.9  &   F439W &  800\nl
G38 & 8296 & Oct 15 1999 &  0:39:47.35 & 40:31:57.9  &   F439W &  800\nl
G44 & 8296 & Oct 30 1999 &  0:40:01.58 & 40:34:14.7  &   F555W &  600\nl
G44 & 8296 & Oct 30 1999 &  0:40:01.58 & 40:34:14.7  &   F555W &  600\nl
G44 & 8296 & Oct 30 1999 &  0:40:01.58 & 40:34:14.7  &   F439W &  800\nl
G44 & 8296 & Oct 30 1999 &  0:40:01.58 & 40:34:14.7  &   F439W &  800\nl
OB78-WR2-FIELD & 6038 & Jan 23 1996 &  0:40:14.10 & 40:37:11.3  &   F555W &  160\nl
OB78-WR2-FIELD & 6038 & Jan 23 1996 &  0:40:14.10 & 40:37:11.3  &   F439W &  600\nl
NGC224-FIELD1 & 6431 & Dec 9 1997 &  0:40:39.54 & 40:33:25.4  &   F439W &  350\tablebreak
NGC224-FIELD1 & 6431 & Dec 9 1997 &  0:40:39.54 & 40:33:25.4  &   F439W &  350\nl
NGC224-FIELD1 & 6431 & Dec 9 1997 &  0:40:39.54 & 40:33:25.4  &   F555W &  260\nl
NGC224-FIELD1 & 6431 & Dec 9 1997 &  0:40:39.54 & 40:33:25.4  &   F555W &  260\nl
NGC224-FIELD1 & 6431 & Dec 9 1997 &  0:40:39.54 & 40:33:25.4  &   F814W &  260\nl
NGC224-FIELD1 & 6431 & Dec 9 1997 &  0:40:39.54 & 40:33:25.4  &   F814W &  260\nl
G76 & 6671 & Jan 11 1999 &  0:40:56.68 & 40:35:28.9  &   F555W & 1200\nl
G76 & 6671 & Jan 11 1999 &  0:40:56.68 & 40:35:28.9  &   F555W & 1400\nl
G76 & 6671 & Jan 11 1999 &  0:40:56.68 & 40:35:28.9  &   F555W & 1300\nl
G76 & 6671 & Jan 11 1999 &  0:40:56.68 & 40:35:28.9  &   F555W & 1400\nl
G76 & 6671 & Jan 11 1999 &  0:40:56.68 & 40:35:28.9  &   F814W & 1300\nl
G76 & 6671 & Jan 11 1999 &  0:40:56.68 & 40:35:28.9  &   F814W & 1400\nl
G76 & 6671 & Jan 11 1999 &  0:40:56.68 & 40:35:28.9  &   F814W & 1300\nl
G76 & 6671 & Jan 11 1999 &  0:40:56.68 & 40:35:28.9  &   F814W & 1400\nl
OB69 & 5998 & Aug 30 1995 &  0:40:59.54 & 41:03:38.4  &   F555W &  160\nl
OB69 & 5998 & Aug 30 1995 &  0:40:59.54 & 41:03:38.4  &   F439W &  500\nl
OB69 & 5998 & Aug 30 1995 &  0:40:59.54 & 41:03:38.4  &   F439W &  300\nl
G87 & 6671 & Aug 16 1999 &  0:41:16.28 & 40:56:12.5  &   F555W & 1200\nl
G87 & 6671 & Aug 16 1999 &  0:41:16.28 & 40:56:12.5  &   F555W & 1400\nl
G87 & 6671 & Aug 16 1999 &  0:41:16.28 & 40:56:12.5  &   F555W & 1300\nl
G87 & 6671 & Aug 16 1999 &  0:41:16.28 & 40:56:12.5  &   F555W & 1400\nl
G87 & 6671 & Aug 16 1999 &  0:41:16.28 & 40:56:12.5  &   F814W & 1300\nl
G87 & 6671 & Aug 16 1999 &  0:41:16.28 & 40:56:12.5  &   F814W & 1287\nl
G87 & 6671 & Aug 16 1999 &  0:41:16.28 & 40:56:12.5  &   F814W & 1300\nl
G87 & 6671 & Aug 16 1999 &  0:41:16.28 & 40:56:12.5  &   F814W & 1400\tablebreak
G94 & 8296 & Oct 30 1999 &  0:41:22.08 & 40:37:06.7  &   F555W &  600\nl
G94 & 8296 & Oct 30 1999 &  0:41:22.08 & 40:37:06.7  &   F555W &  600\nl
G94 & 8296 & Oct 30 1999 &  0:41:22.08 & 40:37:06.7  &   F439W &  800\nl
G94 & 8296 & Oct 30 1999 &  0:41:22.08 & 40:37:06.7  &   F439W &  800\nl
K108 & 5112 & Feb 15 1994 &  0:41:43.30 & 41:34:20.3  &   F814W & 1000\nl
K108 & 5112 & Feb 15 1994 &  0:41:43.30 & 41:34:20.3  &   F814W & 1000\nl
K108 & 5112 & Feb 15 1994 &  0:41:43.30 & 41:34:20.3  &   F555W & 1000\nl
K108 & 5112 & Feb 15 1994 &  0:41:43.30 & 41:34:20.3  &   F555W & 1000\nl
G119 & 6671 & Jun 13 1999 &  0:41:55.58 & 40:47:15.0  &   F555W & 1200\nl
G119 & 6671 & Jun 13 1999 &  0:41:55.58 & 40:47:15.0  &   F555W & 1400\nl
G119 & 6671 & Jun 13 1999 &  0:41:55.58 & 40:47:15.0  &   F555W & 1300\nl
G119 & 6671 & Jun 13 1999 &  0:41:55.58 & 40:47:15.0  &   F555W & 1400\nl
G119 & 6671 & Jun 13 1999 &  0:41:55.58 & 40:47:15.0  &   F814W & 1300\nl
G119 & 6671 & Jun 13 1999 &  0:41:55.58 & 40:47:15.0  &   F814W & 1400\nl
G119 & 6671 & Jun 13 1999 &  0:41:55.58 & 40:47:15.0  &   F814W & 1300\nl
G119 & 6671 & Jun 13 1999 &  0:41:55.58 & 40:47:15.0  &   F814W & 1400\nl
NGC224-FIELD2 & 6431 & Dec 9 1997 &  0:42:05.27 & 40:57:33.9  &   F439W &  350\nl
NGC224-FIELD2 & 6431 & Dec 9 1997 &  0:42:05.27 & 40:57:33.9  &   F439W &  350\nl
NGC224-FIELD2 & 6431 & Dec 9 1997 &  0:42:05.27 & 40:57:33.9  &   F555W &  260\nl
NGC224-FIELD2 & 6431 & Dec 9 1997 &  0:42:05.27 & 40:57:33.9  &   F555W &  260\nl
NGC224-FIELD2 & 6431 & Dec 9 1997 &  0:42:05.27 & 40:57:33.9  &   F814W &  260\nl
NGC224-FIELD2 & 6431 & Dec 9 1997 &  0:42:05.27 & 40:57:33.9  &   F814W &  260\nl
NGC224-DISK & 6636 & Dec 31 1996 &  0:42:18.01 & 40:45:03.7  &   F555W &  600\nl
NGC224-DISK & 6636 & Dec 31 1996 &  0:42:18.01 & 40:45:03.7  &   F555W & 1200\tablebreak
NGC224-DISK & 6636 & Dec 31 1996 &  0:42:18.01 & 40:45:03.7  &   F814W & 1300\nl
NGC224-DISK & 6636 & Dec 31 1996 &  0:42:18.01 & 40:45:03.7  &   F814W & 1200\nl
NGC221-POS2 & 5233 & Oct 22 1994 &  0:43:04.61 & 40:54:33.0  &   F555W &  500\nl
NGC221-POS2 & 5233 & Oct 22 1994 &  0:43:04.61 & 40:54:33.0  &   F555W &  500\nl
NGC221-POS2 & 5233 & Oct 22 1994 &  0:43:04.61 & 40:54:33.0  &   F555W &  500\nl
NGC221-POS2 & 5233 & Oct 22 1994 &  0:43:04.61 & 40:54:33.0  &   F555W &  500\nl
NGC221-POS2 & 5233 & Oct 22 1994 &  0:43:04.61 & 40:54:33.0  &   F814W &  500\nl
NGC221-POS2 & 5233 & Oct 22 1994 &  0:43:04.61 & 40:54:33.0  &   F814W &  500\nl
NGC221-POS2 & 5233 & Oct 22 1994 &  0:43:04.61 & 40:54:33.0  &   F814W &  500\nl
NGC221-POS2 & 5233 & Oct 22 1994 &  0:43:04.61 & 40:54:33.0  &   F814W &  500\nl
INNER & 6859 & Aug 14 1997 &  0:44:23.74 & 41:45:16.2  &   F555W &  300\nl
INNER & 6859 & Aug 14 1997 &  0:44:23.74 & 41:45:16.2  &   F555W & 1000\nl
INNER & 6859 & Aug 14 1997 &  0:44:23.74 & 41:45:16.2  &   F555W & 1100\nl
INNER & 6859 & Aug 14 1997 &  0:44:23.74 & 41:45:16.2  &   F814W &  200\nl
INNER & 6859 & Aug 14 1997 &  0:44:23.74 & 41:45:16.2  &   F814W & 1000\nl
INNER & 6859 & Aug 14 1997 &  0:44:23.74 & 41:45:16.2  &   F814W & 1300\nl
G287 & 6671 & Sep 26 1999 &  0:44:42.45 & 41:44:24.2  &   F555W & 1200\nl
G287 & 6671 & Sep 26 1999 &  0:44:42.45 & 41:44:24.2  &   F555W & 1400\nl
G287 & 6671 & Sep 26 1999 &  0:44:42.45 & 41:44:24.2  &   F555W & 1300\nl
G287 & 6671 & Sep 26 1999 &  0:44:42.45 & 41:44:24.2  &   F555W & 1400\nl
G287 & 6671 & Sep 26 1999 &  0:44:42.45 & 41:44:24.2  &   F814W & 1300\nl
G287 & 6671 & Sep 26 1999 &  0:44:42.45 & 41:44:24.2  &   F814W & 1400\nl
G287 & 6671 & Sep 26 1999 &  0:44:42.45 & 41:44:24.2  &   F814W & 1300\nl
G287 & 6671 & Sep 26 1999 &  0:44:42.45 & 41:44:24.2  &   F814W & 1400\tablebreak
G11 & 6671 & Sep 25 1999 &  0:44:42.52 & 41:44:24.1  &   F555W & 1200\nl
G11 & 6671 & Sep 25 1999 &  0:44:42.52 & 41:44:24.1  &   F555W & 1400\nl
G11 & 6671 & Sep 25 1999 &  0:44:42.52 & 41:44:24.1  &   F555W & 1300\nl
G11 & 6671 & Sep 25 1999 &  0:44:42.52 & 41:44:24.1  &   F555W & 1400\nl
G11 & 6671 & Sep 25 1999 &  0:44:42.52 & 41:44:24.1  &   F814W & 1300\nl
G11 & 6671 & Sep 25 1999 &  0:44:42.52 & 41:44:24.1  &   F814W & 1400\nl
G11 & 6671 & Sep 25 1999 &  0:44:42.52 & 41:44:24.1  &   F814W & 1300\nl
G11 & 6671 & Sep 25 1999 &  0:44:42.52 & 41:44:24.1  &   F814W & 1400\nl
NGC224-VDB-OB042-POS01 & 5911 & Oct 3 1995 &  0:44:44.23 & 41:27:33.8  &   F439W &  160\nl
NGC224-VDB-OB042-POS01 & 5911 & Oct 3 1995 &  0:44:44.23 & 41:27:33.8  &   F555W &  140\nl
G213 & 8296 & Oct 31 1999 &  0:44:46.19 & 41:51:33.3  &   F555W &  600\nl
G213 & 8296 & Oct 31 1999 &  0:44:46.19 & 41:51:33.3  &   F555W &  600\nl
G213 & 8296 & Oct 31 1999 &  0:44:46.19 & 41:51:33.3  &   F439W &  800\nl
G213 & 8296 & Oct 31 1999 &  0:44:46.19 & 41:51:33.3  &   F439W &  800\nl
NGC224-VDB-OB042-POS02 & 5911 & Oct 8 1995 &  0:44:49.34 & 41:28:59.0  &   F439W &  160\nl
NGC224-VDB-OB042-POS02 & 5911 & Oct 8 1995 &  0:44:49.34 & 41:28:59.0  &   F555W &  140\nl
G272 & 5420 & Jan 22 1995 &  0:44:50.61 & 41:19:11.1  &   F555W & 1500\nl
G272 & 5420 & Jan 22 1995 &  0:44:50.61 & 41:19:11.1  &   F555W & 2300\nl
G272 & 5420 & Jan 22 1995 &  0:44:50.61 & 41:19:11.1  &   F814W & 2300\nl
G272 & 5420 & Jan 22 1995 &  0:44:50.61 & 41:19:11.1  &   F814W & 2300\nl
G272 & 5420 & Jan 22 1995 &  0:44:50.61 & 41:19:11.1  &   F814W & 2300\nl
G272 & 5420 & Jan 22 1995 &  0:44:50.61 & 41:19:11.1  &   F814W & 2300\nl
G272 & 5420 & Jan 22 1995 &  0:44:50.61 & 41:19:11.1  &   F814W & 1600\nl
OB48-307-FIELD & 6038 & Jan 1 1996 &  0:44:51.22 & 41:30:03.7  &   F555W &  160\tablebreak
OB48-307-FIELD & 6038 & Jan 1 1996 &  0:44:51.22 & 41:30:03.7  &   F439W &  600\nl
NGC224-VDB-OB042-POS03 & 5911 & Oct 4 1995 &  0:44:57.63 & 41:30:51.6  &   F439W &  160\nl
NGC224-VDB-OB042-POS03 & 5911 & Oct 4 1995 &  0:44:57.63 & 41:30:51.6  &   F555W &  140\nl
NGC224-VDB-OB048-POS01 & 5911 & Oct 15 1995 &  0:45:09.25 & 41:34:30.7  &   F439W &  160\nl
NGC224-VDB-OB048-POS01 & 5911 & Oct 15 1995 &  0:45:09.25 & 41:34:30.7  &   F555W &  140\nl
NGC224-VDB-OB048-POS02 & 5911 & Oct 15 1995 &  0:45:11.95 & 41:36:57.0  &   F439W &  160\nl
NGC224-VDB-OB048-POS02 & 5911 & Oct 15 1995 &  0:45:11.95 & 41:36:57.0  &   F555W &  140\nl
G322 & 6671 & Jan 10 1999 &  0:46:24.56 & 42:01:38.6  &   F555W & 1200\nl
G322 & 6671 & Jan 10 1999 &  0:46:24.56 & 42:01:38.6  &   F555W & 1400\nl
G322 & 6671 & Jan 10 1999 &  0:46:24.56 & 42:01:38.6  &   F555W & 1300\nl
G322 & 6671 & Jan 10 1999 &  0:46:24.56 & 42:01:38.6  &   F555W & 1400\nl
G322 & 6671 & Jan 10 1999 &  0:46:24.56 & 42:01:38.6  &   F814W & 1300\nl
G322 & 6671 & Jan 10 1999 &  0:46:24.56 & 42:01:38.6  &   F814W & 1400\nl
G322 & 6671 & Jan 10 1999 &  0:46:24.56 & 42:01:38.6  &   F814W & 1300\nl
G322 & 6671 & Jan 10 1999 &  0:46:24.56 & 42:01:38.6  &   F814W & 1400\nl
\enddata
\end{deluxetable}
\end{landscape}

\clearpage
\begin{deluxetable}{ccc}
\small
\tablecaption{Distances and mean reddening values determined by the MATCH softwarefor each field.}
\tablehead{
\colhead{\bf{Field}} &
\colhead{\bf{m-M}} &
\colhead{\bf{mean A$_V$}} 
}
\startdata
OUTER & 24.45$\pm$0.050 & 0.400$\pm$0.100\nl
G33 & 24.471$\pm$0.063 & 0.367$\pm$0.103\nl
G38 & 24.450$\pm$0.038 & 0.700$\pm$0.076\nl
G44 & 24.450$\pm$0.038 & 0.700$\pm$0.076\nl
OB78-WR2-FIELD & 24.456$\pm$0.039 & 0.712$\pm$0.078\nl
NGC224-FIELD1 B\&V& 24.465$\pm$0.032 & 1.000$\pm$0.141\nl
NGC224-FIELD1 V\&I& 24.450$\pm$0.071 & 0.770$\pm$0.064\nl
G76 & 24.430$\pm$0.060 & 0.500$\pm$0.077\nl
OB69 & 24.450$\pm$0.038 & 1.071$\pm$0.103\nl
G87 & 24.485$\pm$0.055 & 0.690$\pm$0.094\nl
G94 & 24.443$\pm$0.042 & 0.771$\pm$0.070\nl
K108 & 24.435$\pm$0.063 & 0.160$\pm$0.066\nl
G119 & 24.465$\pm$0.063 & 0.220$\pm$0.075\nl
NGC224-FIELD2 B\&V & 24.460$\pm$0.037 & 0.670$\pm$0.110\nl
NGC224-FIELD2 V\&I & 24.480$\pm$0.064 & 0.520$\pm$0.098\nl
NGC224-DISK & 24.436$\pm$0.061 & 0.191$\pm$0.079\nl
NGC221-POS2 & 24.460$\pm$0.062 & 0.200$\pm$0.077\nl
INNER & 24.505$\pm$0.042 & 0.280$\pm$0.098\nl
G287 & 24.500$\pm$0.039 & 0.320$\pm$0.098\nl
G11 & 24.490$\pm$0.049 & 0.310$\pm$0.083\nl
NGC224-VDB-OB042-POS01 & 24.464$\pm$0.035 & 0.914$\pm$0.136\nl
G213 & 24.443$\pm$0.042 & 0.771$\pm$0.070\nl
NGC224-VDB-OB042-POS02 & 24.460$\pm$0.037 & 0.840$\pm$0.049\nl
G272 & 24.450$\pm$0.071 & 0.190$\pm$0.070\nl
OB48-307-FIELD & 24.462$\pm$0.033 & 0.862$\pm$0.099\nl
NGC224-VDB-OB042-POS03 & 24.475$\pm$0.025 & 1.000$\pm$0.082\nl
NGC224-VDB-OB048-POS01 & 24.480$\pm$0.024 & 1.060$\pm$0.136\nl
NGC224-VDB-OB048-POS02 & 24.475$\pm$0.025 & 0.700$\pm$0.082\nl
G322 & 24.445$\pm$0.062 & 0.291$\pm$0.079\nl
\enddata
\end{deluxetable} 

\begin{figure}
\centerline{\psfig{file=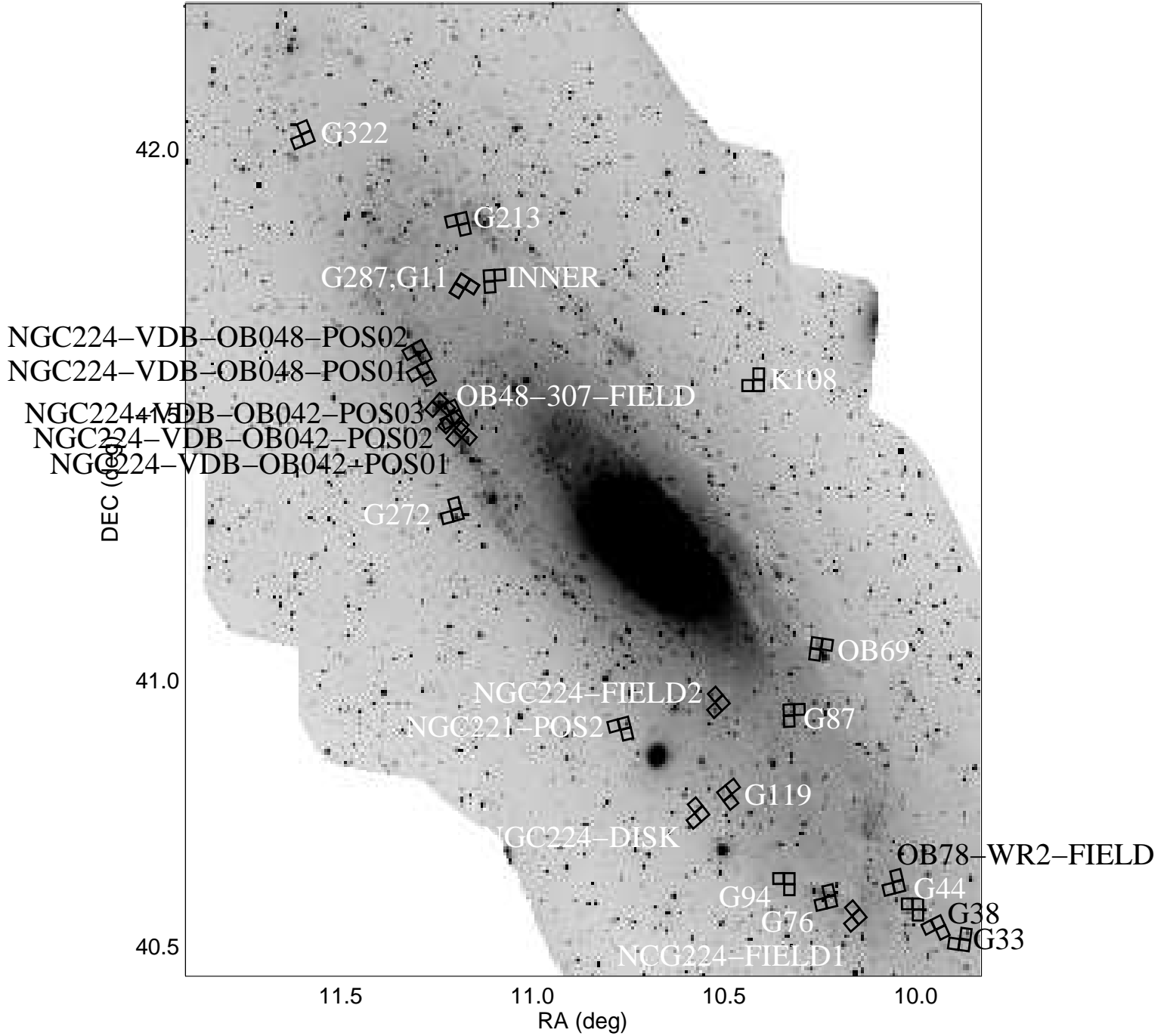 ,height=8.0in,angle=270}}
\caption{Positions of the HST fields taken from the HST archive.
Exposure times and filters observed are given in Table 1.}
\end{figure} 


\begin{figure}
\figurenum{2}
\centerline{\psfig{file=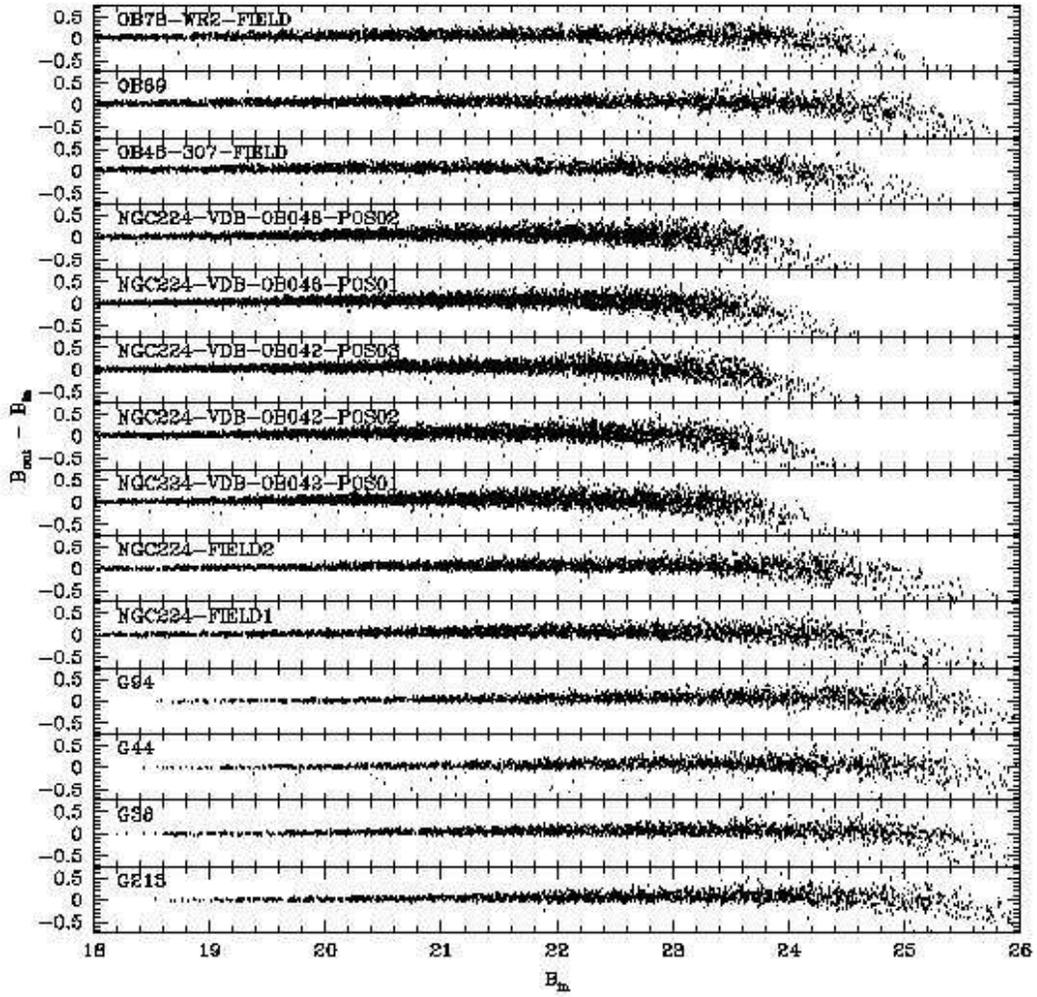,height=8.0in,angle=0}}
\caption{Photometric errors from the artificial star test results
performed on each field. 
(a)  The B band errors.}  
\end{figure}

\begin{figure}
\figurenum{2} 
\centerline{\psfig{file=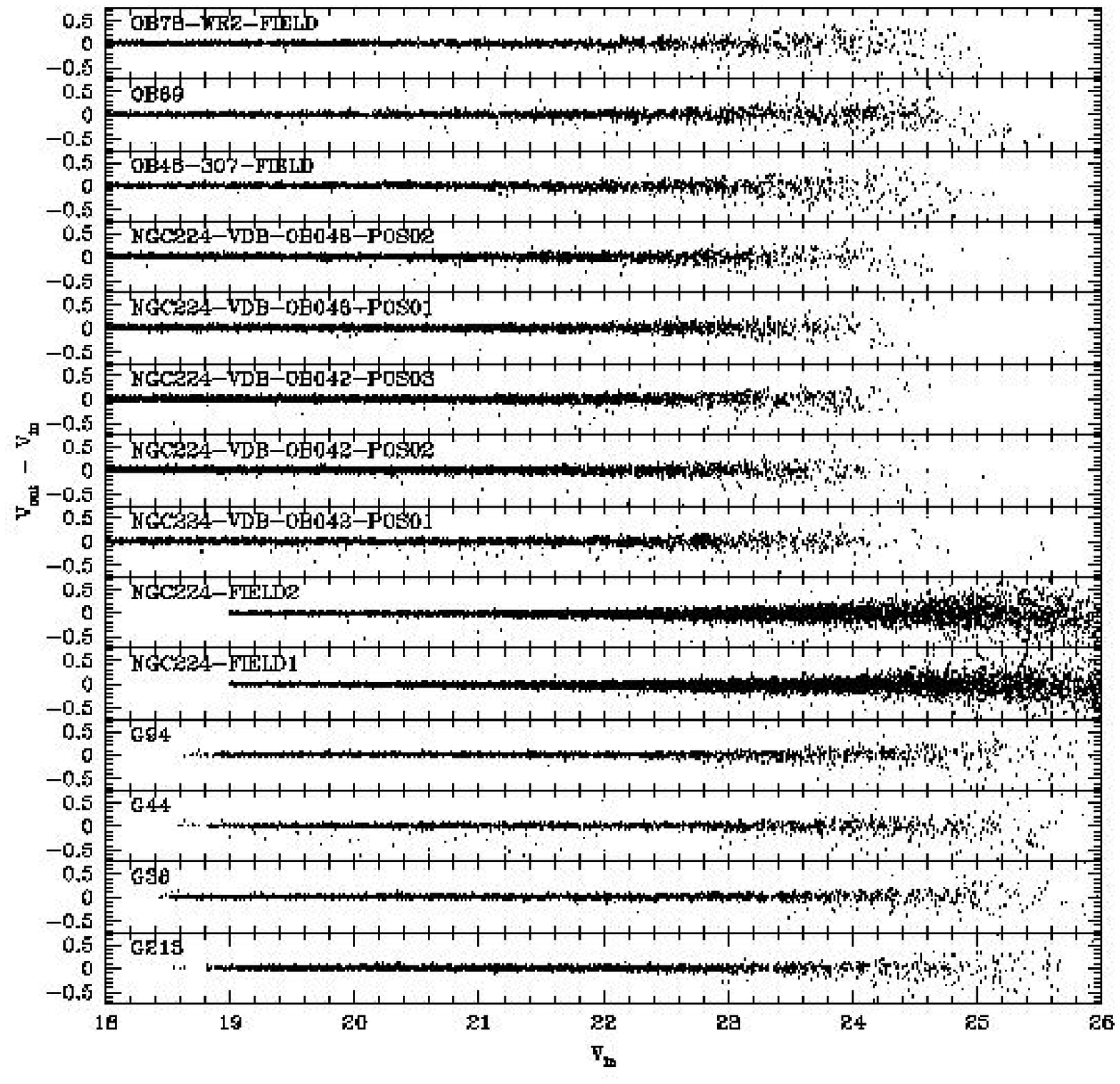,height=8.0in,angle=0}}
\caption{(b)  The V band errors.}  
\end{figure}
 
\begin{figure}
\figurenum{2}
\centerline{\psfig{file=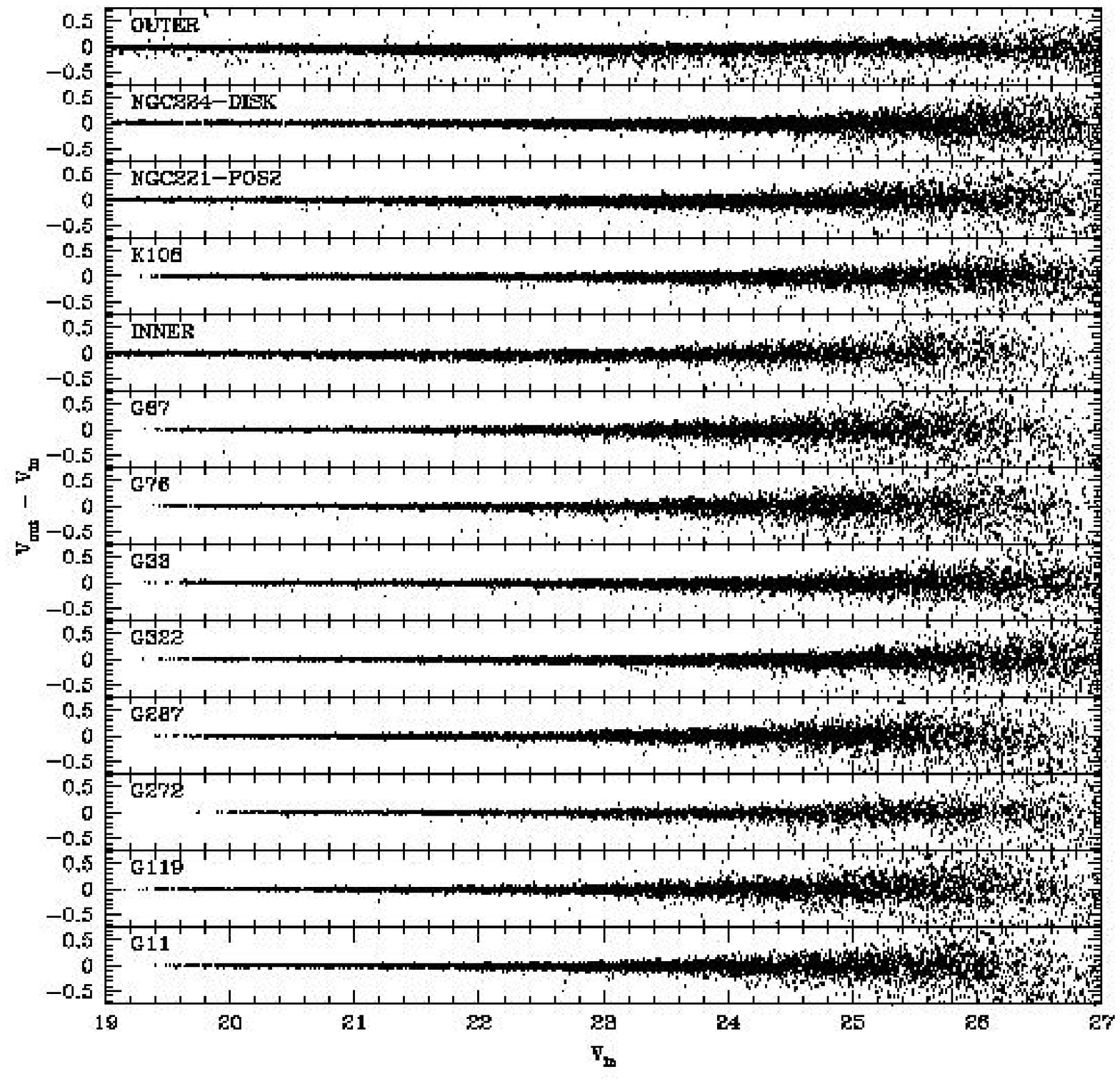,height=8.0in,angle=0}}
\caption{(b)  continued.}  
\end{figure} 

\begin{figure}
\figurenum{2}
\centerline{\psfig{file=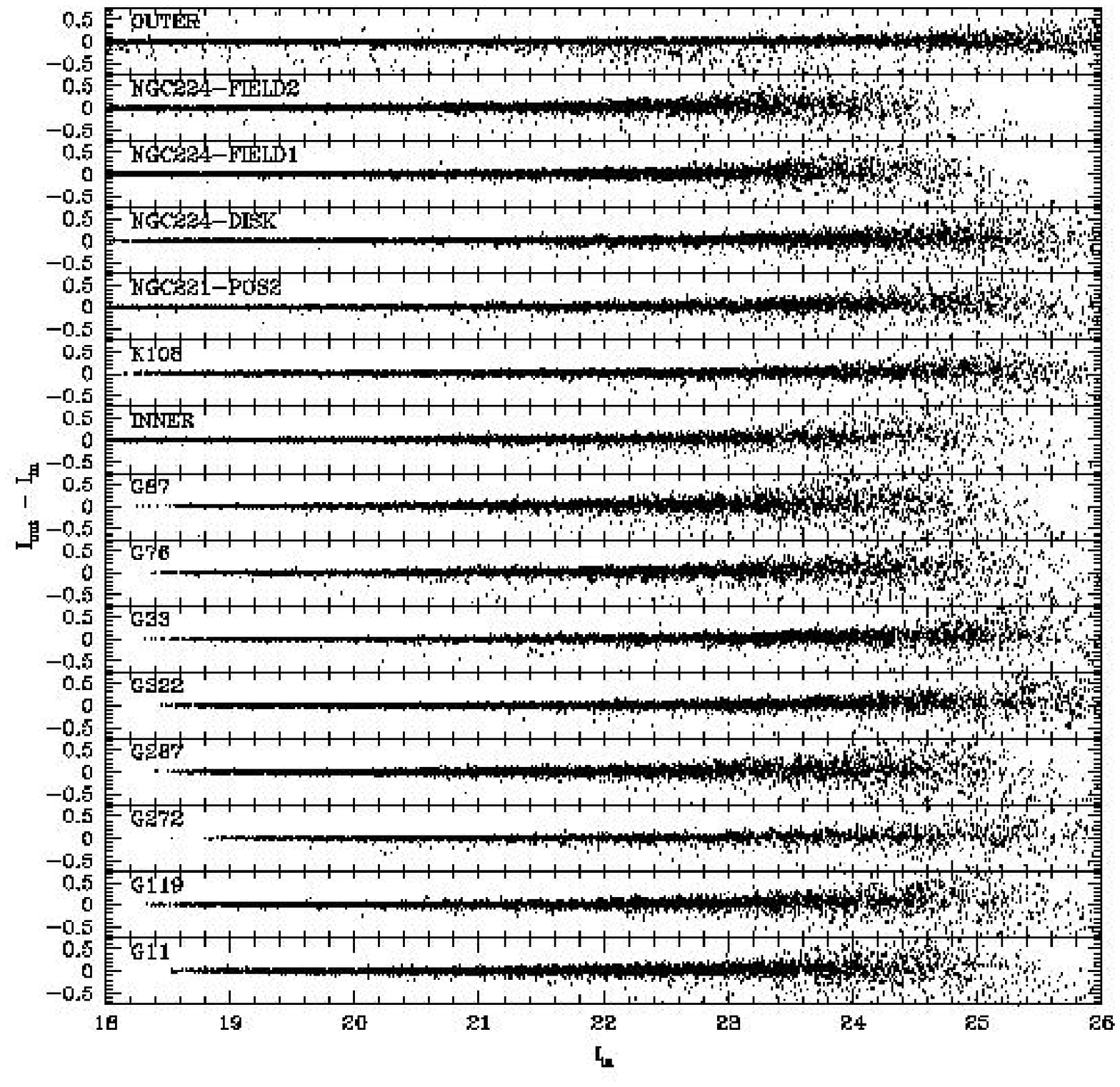,height=7.0in,angle=0}} 
\caption{(c)  The I band errors.}
\end{figure}

\clearpage 
\oddsidemargin -1.5cm
\begin{landscape}
\begin{figure}
\figurenum{3}
\centerline{\psfig{file=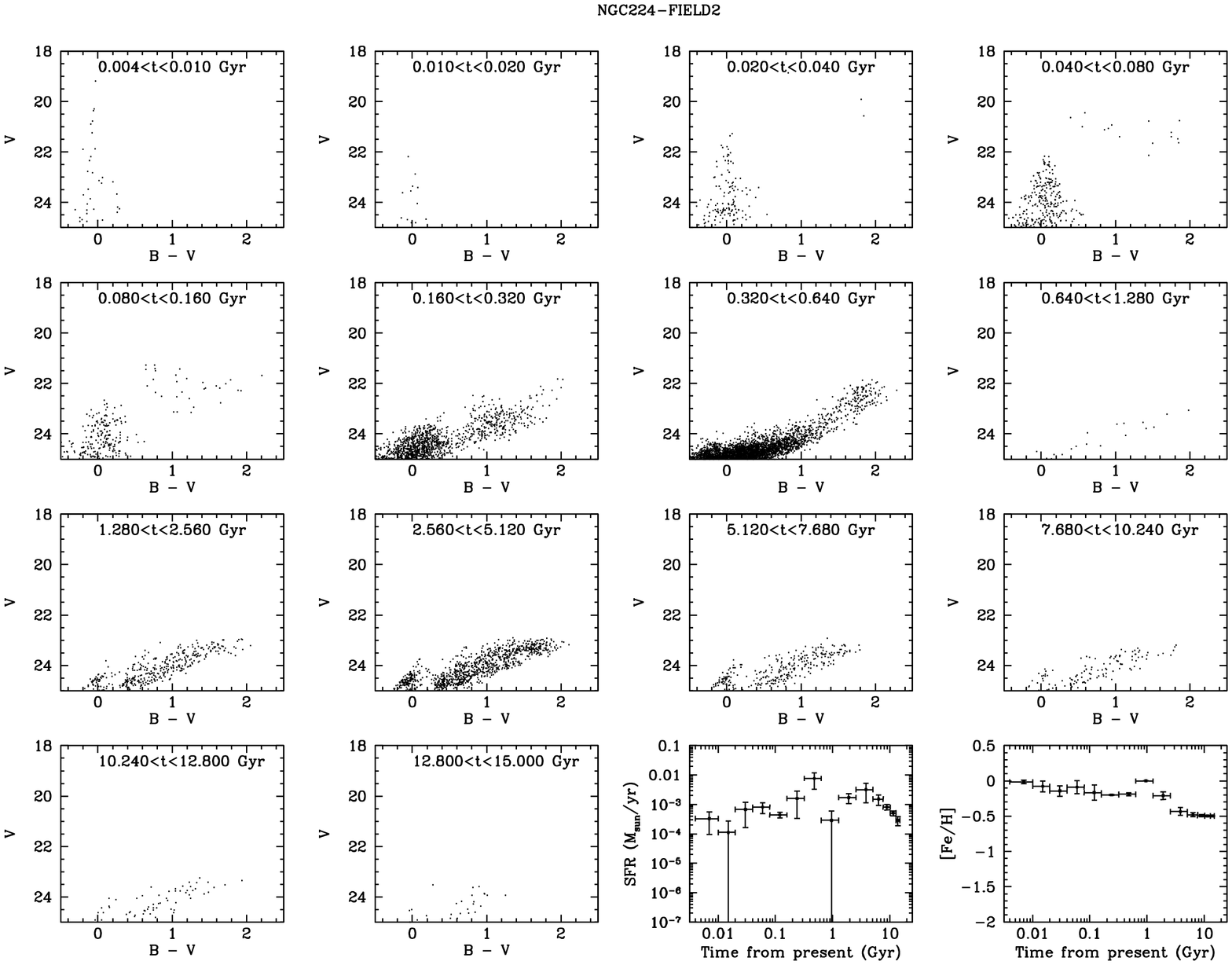,height=7.0in,angle=270}} 
\caption{(a)  The stars formed during each timestep are shown in the 14 V, B-V CMDs labeled by the time periods they represent.  The last two panels show the star formation and metal abundance history which created the CMDs.} 
\end{figure}
\end{landscape} 

\begin{landscape}
\begin{figure}
\figurenum{3}
\centerline{\psfig{file=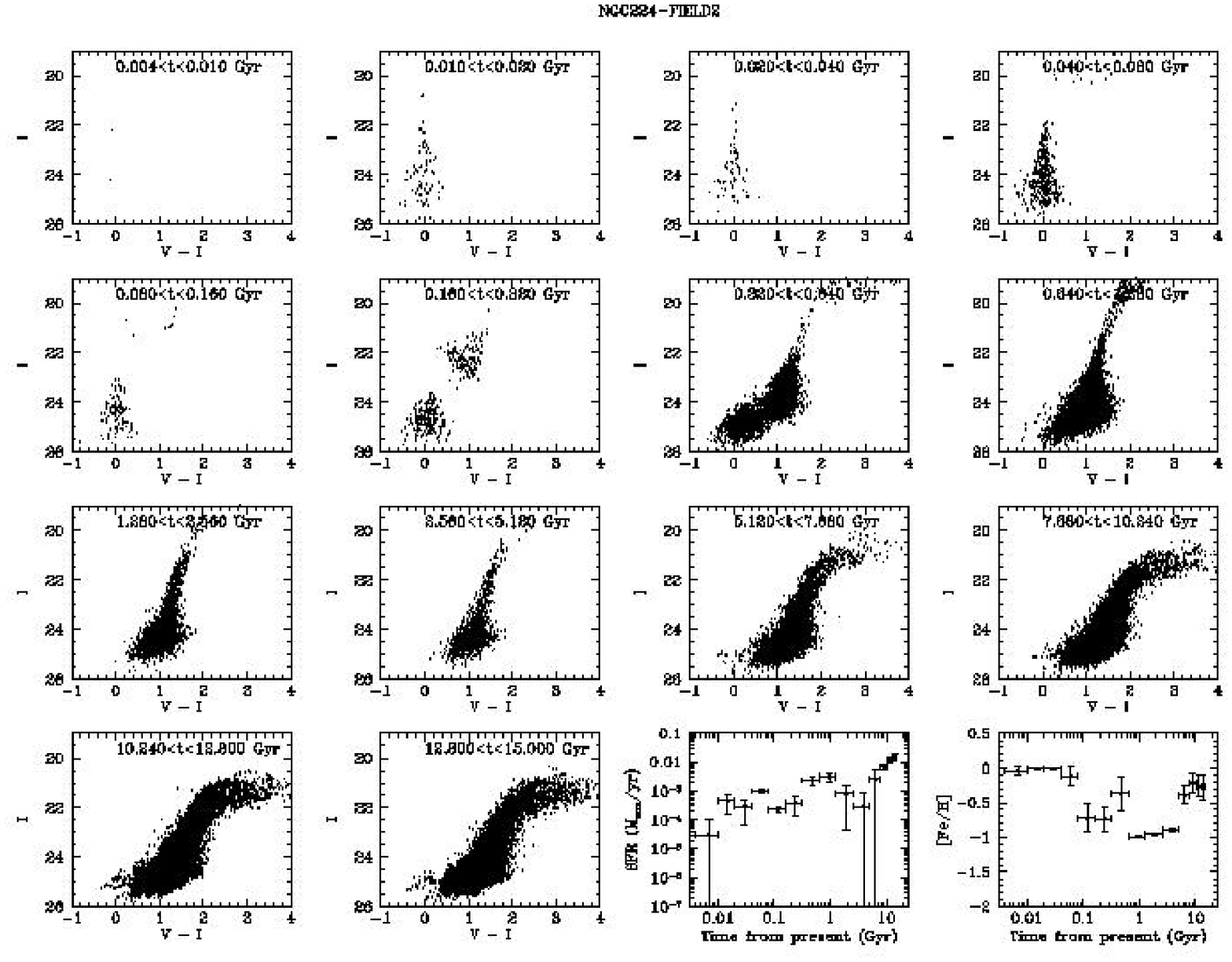,height=7.0in,angle=270}} 
\caption{(b)  The stars formed during each timestep are shown in the 14 I, V-I CMDs labeled by the time periods they represent.  The last two panels show the star formation and metal abundance history which created the CMDs.} 
\end{figure}
\end{landscape} 
 
\oddsidemargin 0cm

\begin{figure}
\figurenum{4}
\centerline{\psfig{file=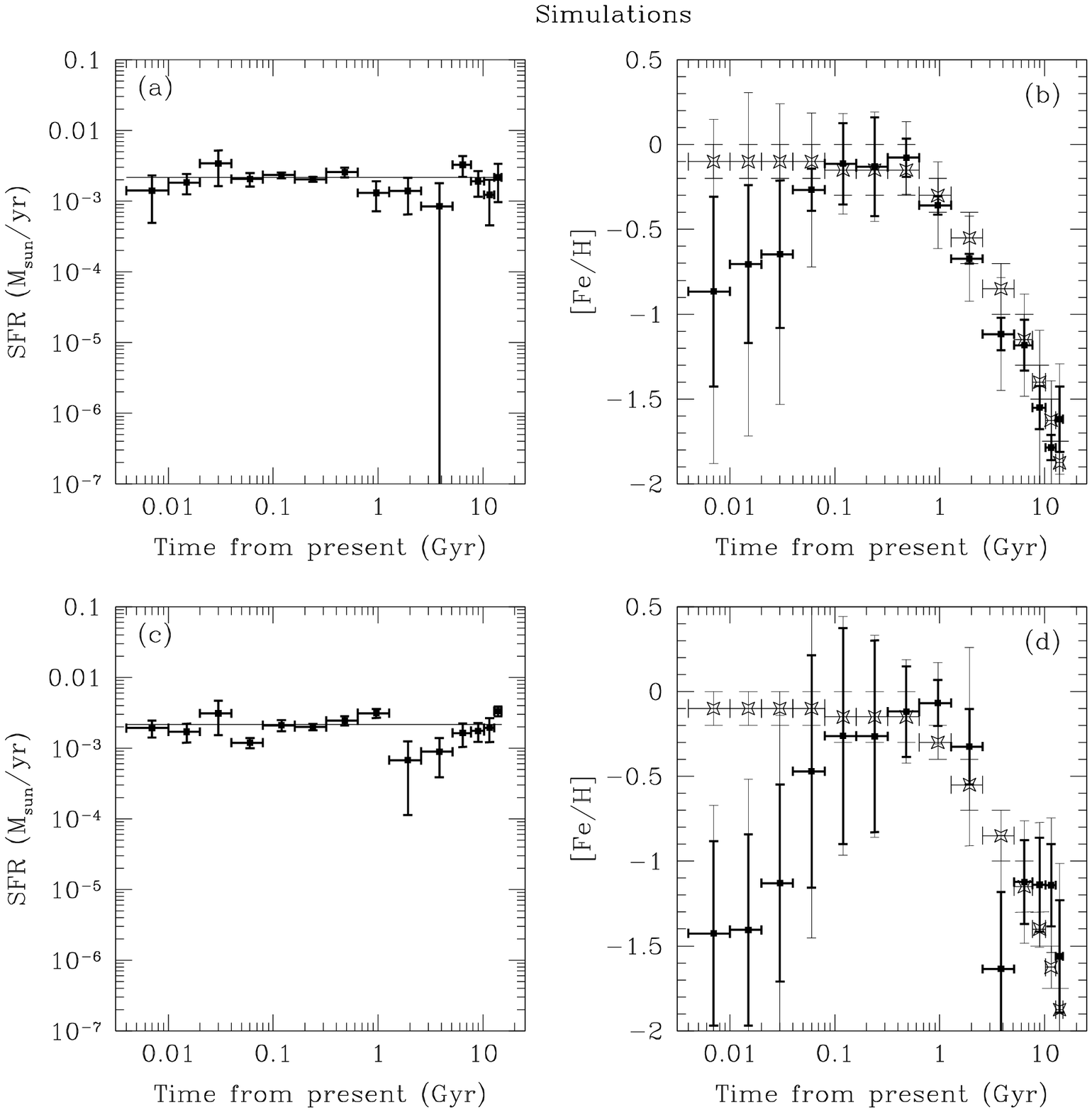,height=6.0in,angle=0}} 

\caption{The results from the artificial field experiment are
shown. (a) Points with error bars show the SFH determined for the
constant star formation rate (shown with the solid line) put into the
analysis routine. (b) Solid square points with error bars show the
abundance history determined by the analysis routine. Heavy error bars
mark the metallicity range for the time period, and the light error
bars show how the measured errors of the mean metal abundance for the
time period could shift the metallicity range.  The input abundance
history is shown by the hollow stars with midweight error bars.  (c)
Same as (a), but the stars in the field were given unique reddening
values, $A_V$ ranging from +0.3 to -0.3 from the mean value.  (d) Same
as (b), but the stars in the field were given unique reddening values,
$A_V$ ranging from +0.3 to -0.3 from the mean value. }

\end{figure} 

\begin{figure}
\figurenum{5}
\centerline{\psfig{file=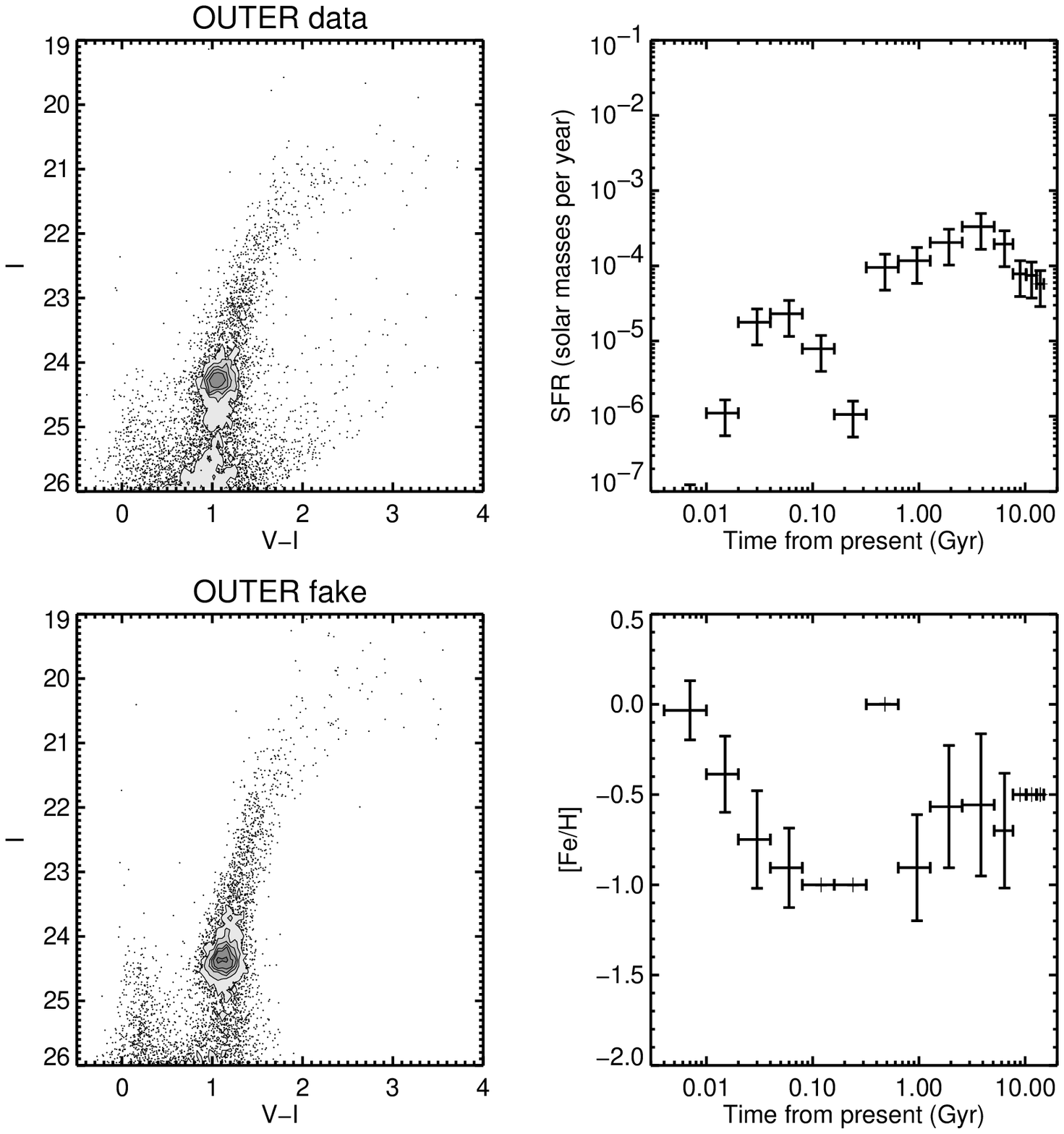,height=6.0in,angle=0}} 

\caption{The SFH and chemical evolution history of some of the M31
fields.  All fields are shown in the electronic version of the
publication.  The upper left panel shows the observed CMD.  The upper
right panel shows the best fitting SFH.  The lower right panel gives
the best fitting chemical evolution; heavy error bars mark the
metallicity range for the time period and light error bars include the
effects of errors in the measured mean metal abundance.  The lower
left panel shows the CMD created from stellar evolution models, using
the SFH shown, assuming a Salpeter IMF and applying errors and
completeness from artificial star tests.  Contours show the stellar
density in areas where the data points would otherwise saturate the
graph.  
\newline 
\smallskip 
(a) The OUTER field, not shown in Figure 1 due to its large galactocentric distance.}

\end{figure}

\begin{figure}
\figurenum{5}
\centerline{\psfig{file=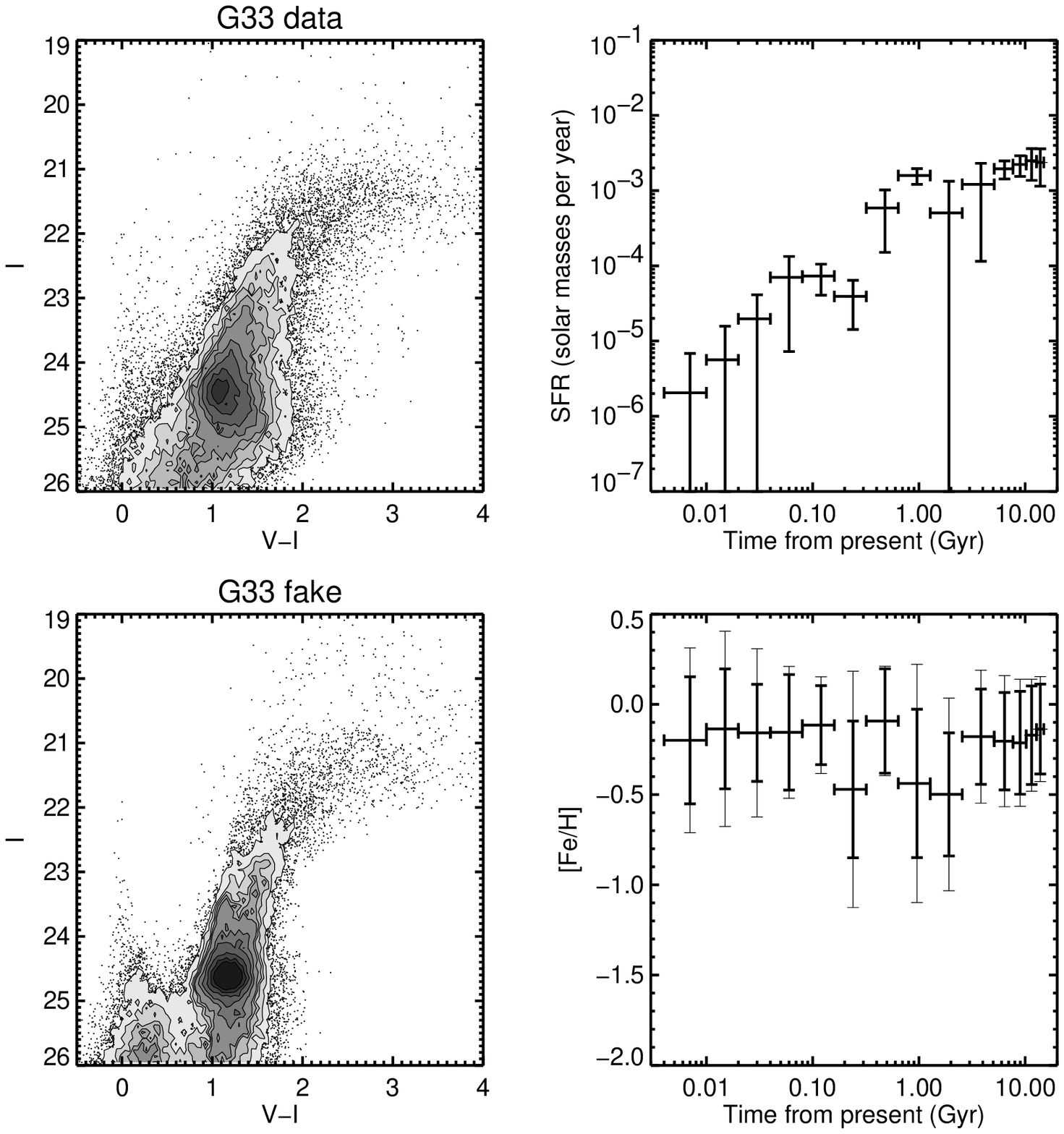,height=6.0in,angle=0}} 
\caption{(b)  The G33 field, which shows very little recent star formation despite its location within the disk.}
\end{figure}     

\begin{figure}
\figurenum{5}
\centerline{\psfig{file=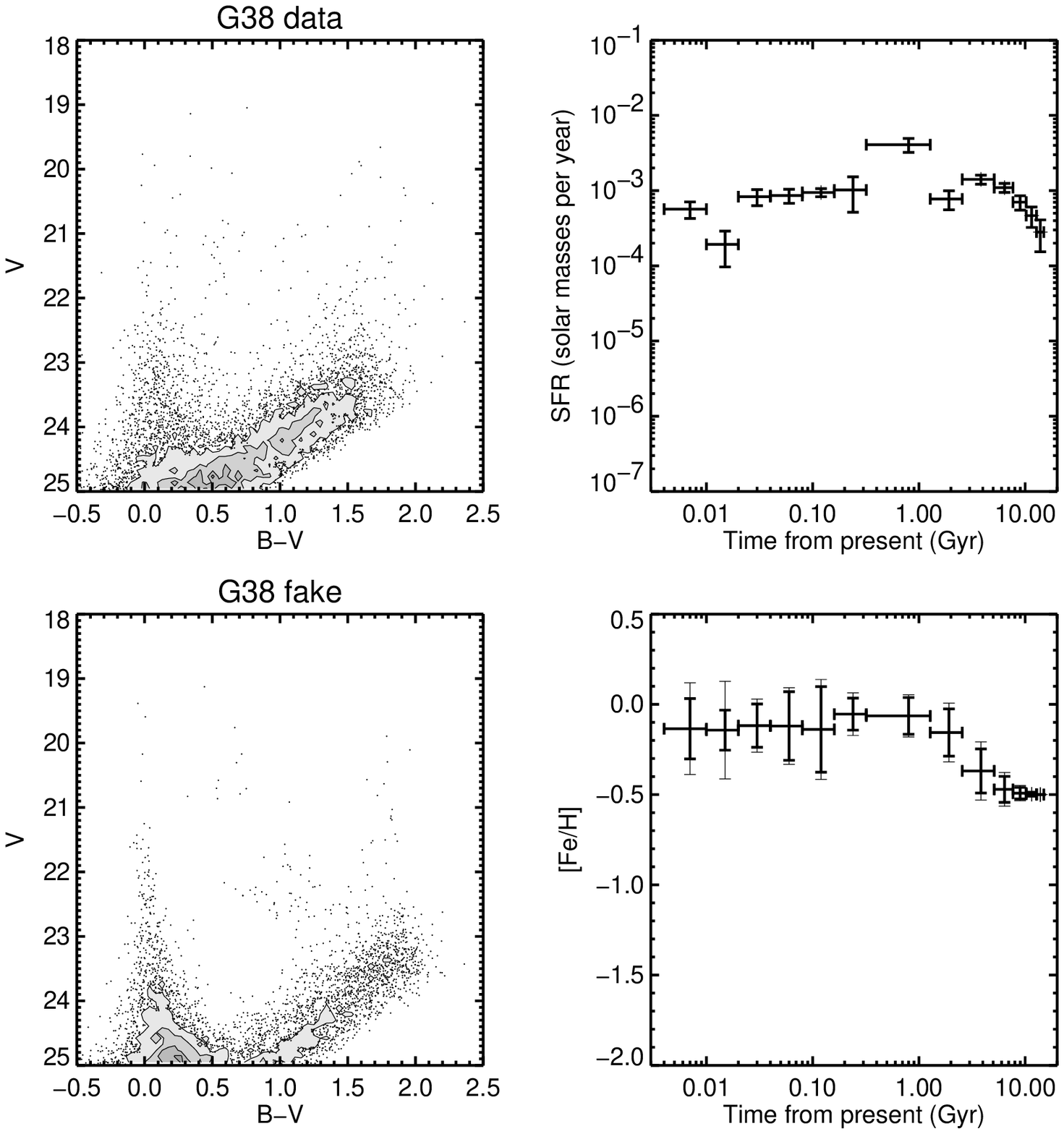,height=6.0in,angle=0}} 
\caption{(c)  The G38 field.}
\end{figure}     

\begin{figure}
\figurenum{5}
\centerline{\psfig{file=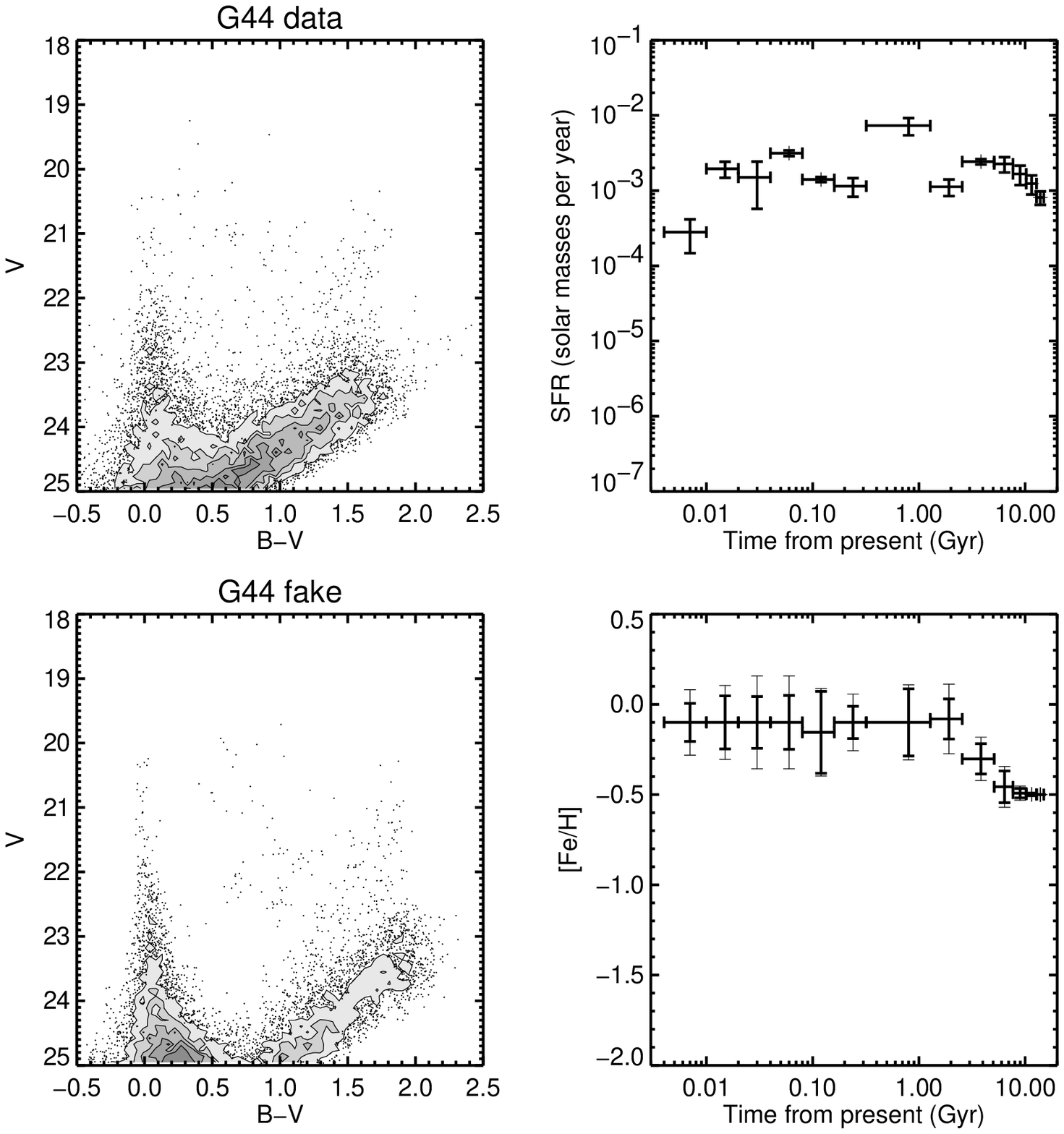,height=6.0in,angle=0}} 
\caption{(d)  The G44 field.}
\end{figure}

\begin{figure}
\figurenum{5}
\centerline{\psfig{file=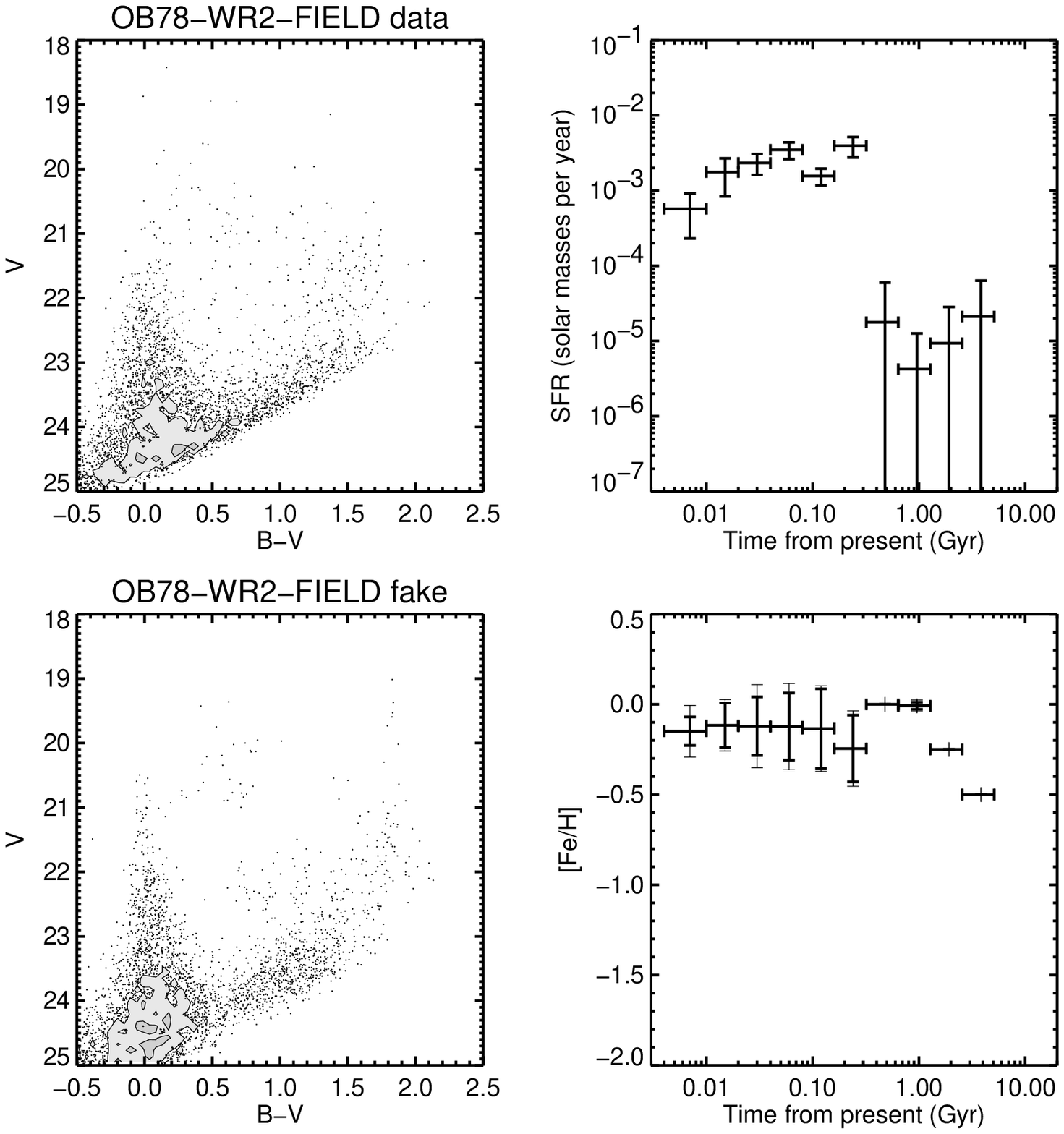,height=6.0in,angle=0}} 
\caption{(e)  The OB78-WR2-FIELD field, only observed in B and V with short exposures, is only sensitive to the recent SFH which suggests a slight decline in activity over the past 100 Myr.}
\end{figure}  

\begin{figure}
\figurenum{5}
\centerline{\psfig{file=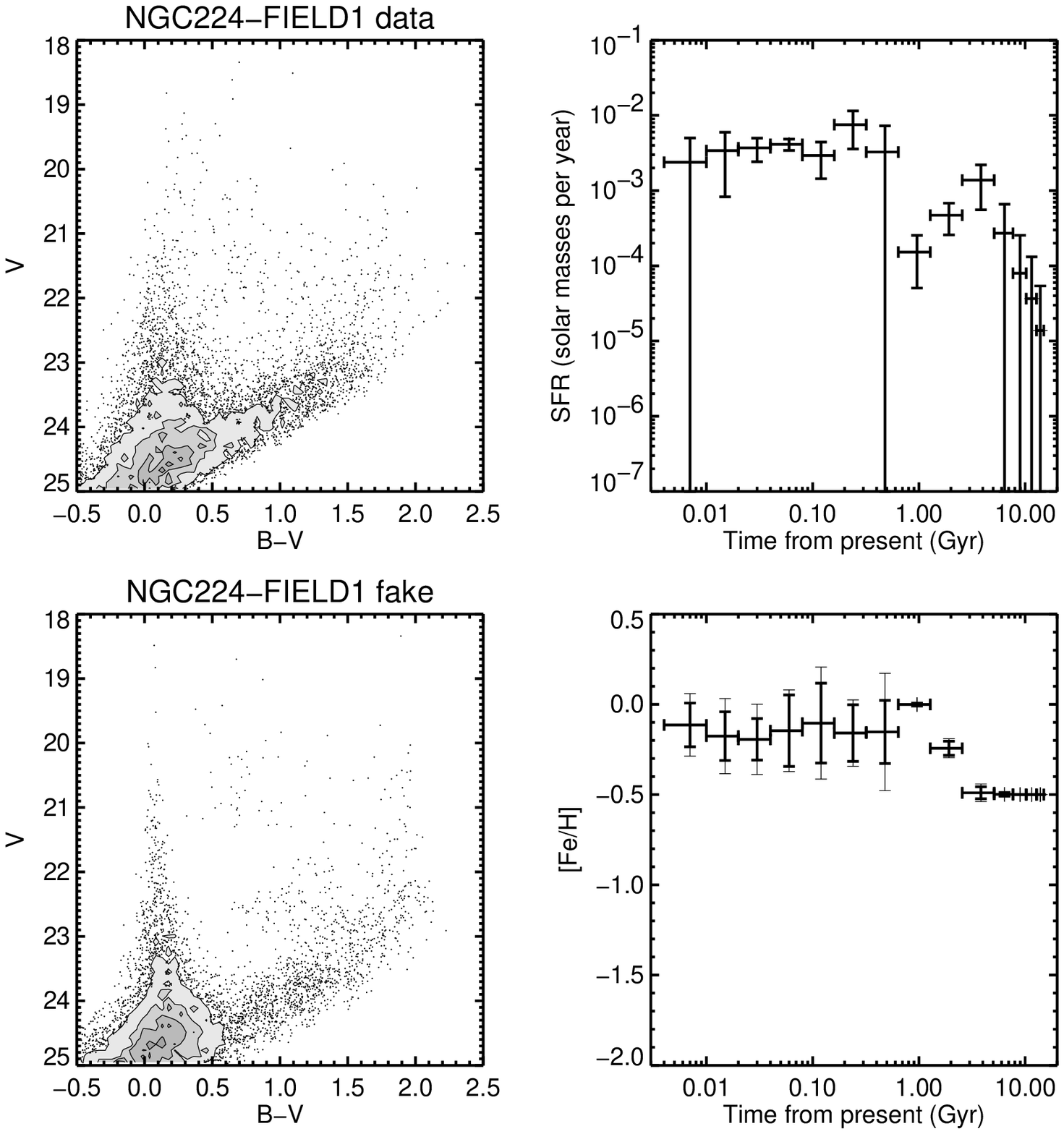,height=6.0in,angle=0}} 
\caption{(f)  The NGC224-FIELD1 field, which was observed in B, V and I.}
\end{figure}      

\begin{figure}
\figurenum{5}
\centerline{\psfig{file=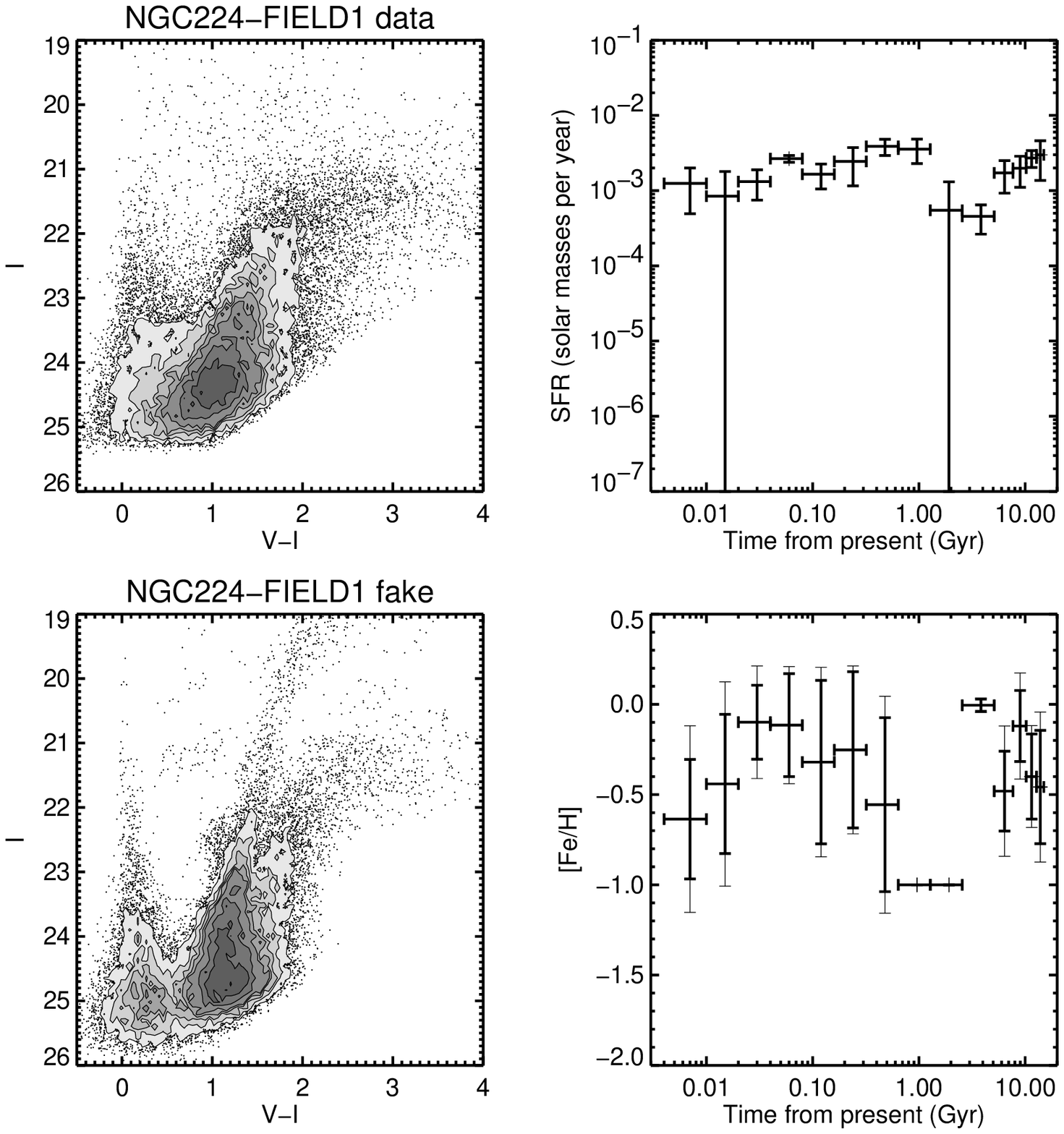,height=6.0in,angle=0}} 
\caption{(g)  The NGC224-FIELD1 field, which was observed in B, V and I.}
\end{figure}  

\begin{figure}
\figurenum{5}
\centerline{\psfig{file=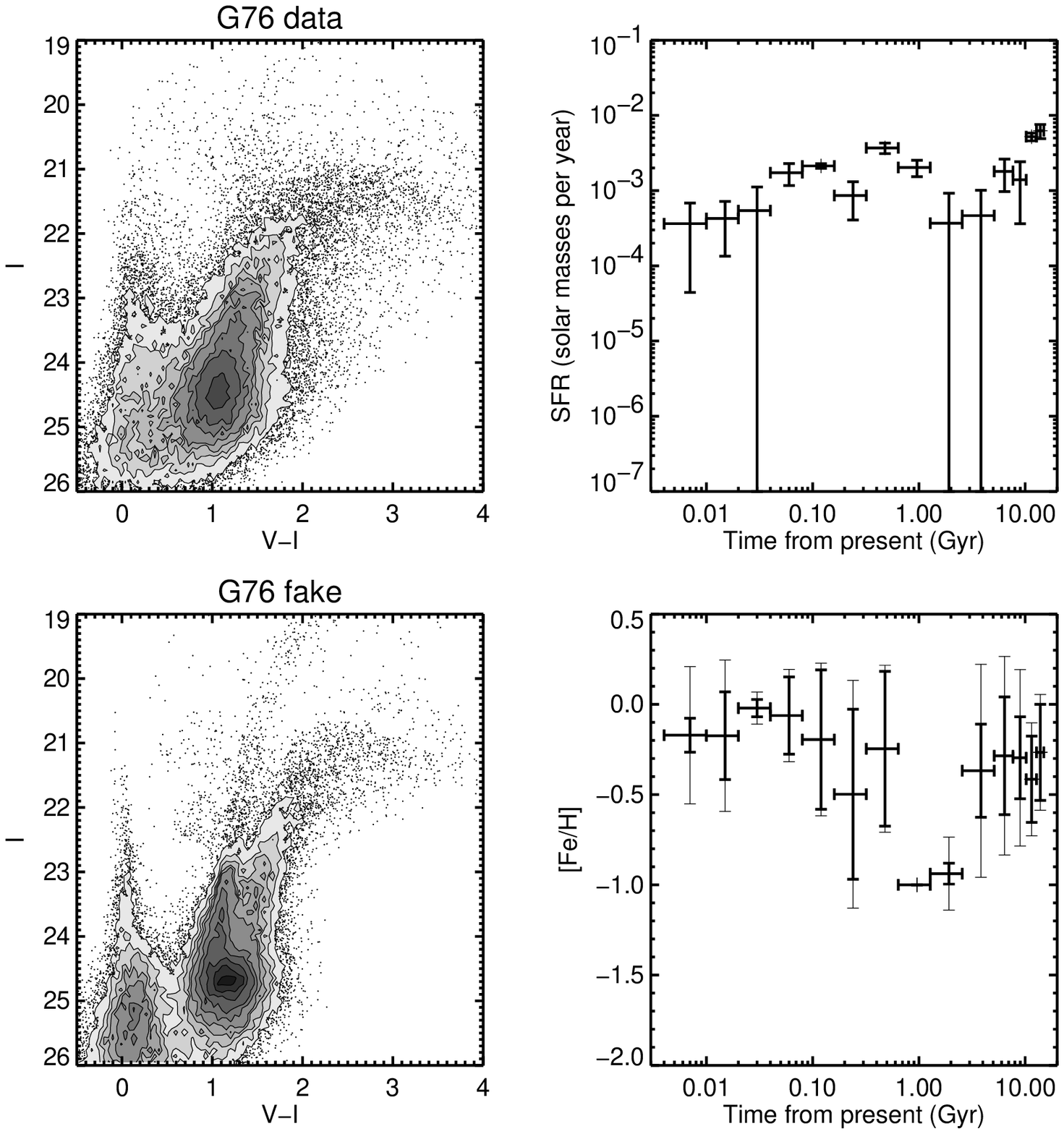,height=6.0in,angle=0}} 
\caption{(h)  The G76 field.}
\end{figure}

\begin{figure}
\figurenum{5}
\centerline{\psfig{file=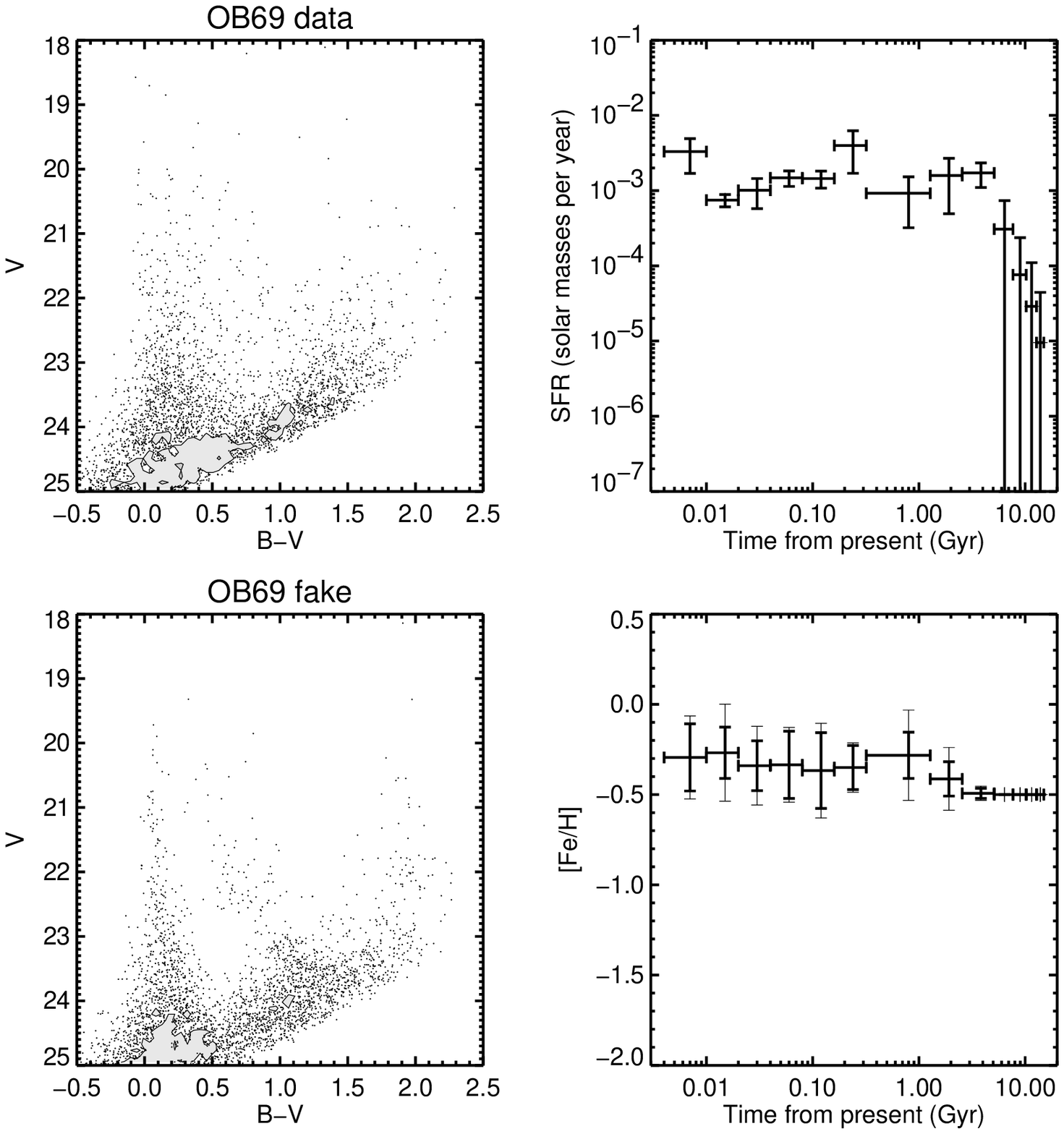,height=6.0in,angle=0}} 
\caption{(i)  The OB69 field.}
\end{figure}

\clearpage

\begin{figure}
\figurenum{5}
\centerline{\psfig{file=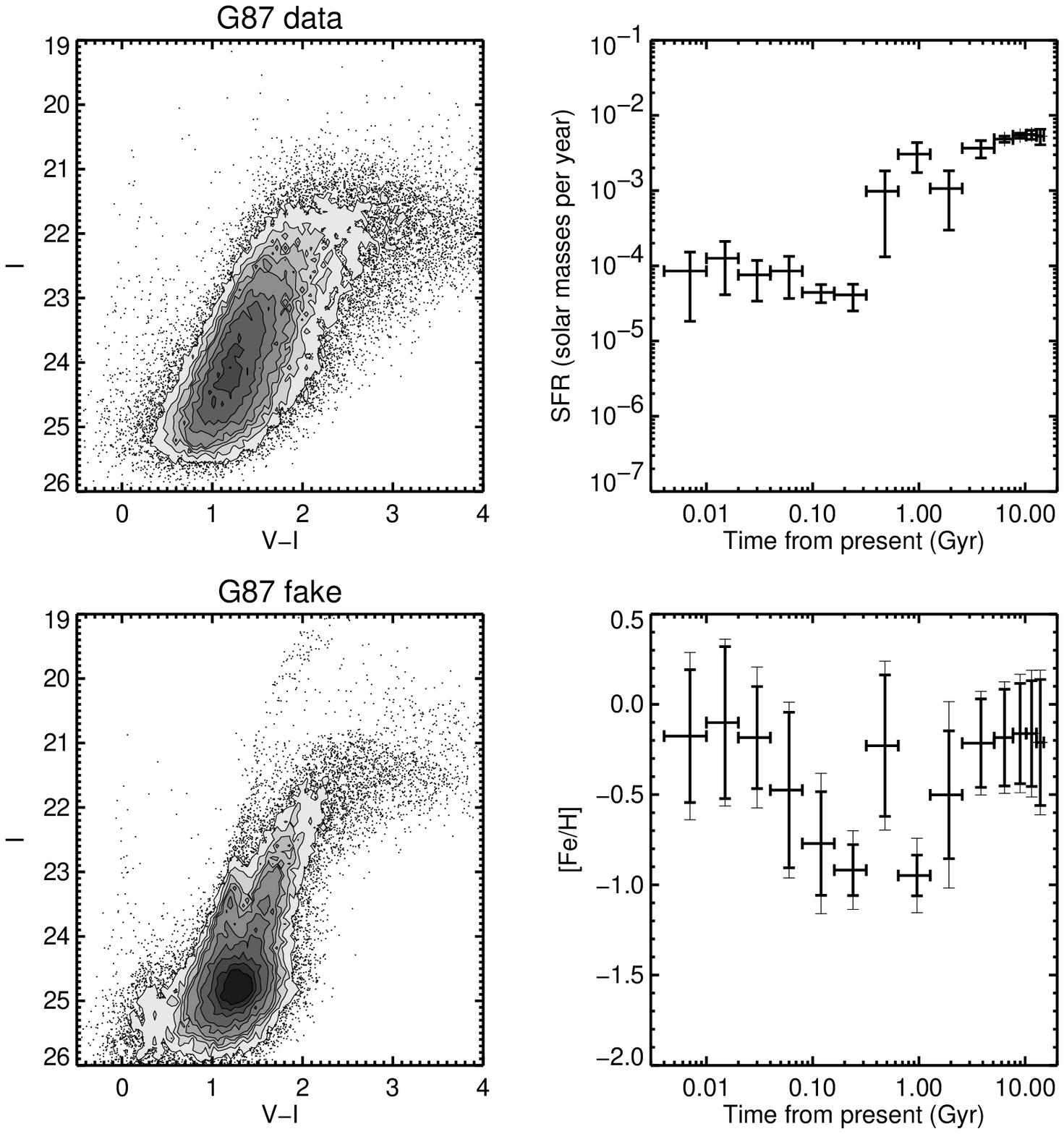,height=6.0in,angle=0}} 
\caption{(j)  The G87 field.}
\end{figure}    

\begin{figure}
\figurenum{5}
\centerline{\psfig{file=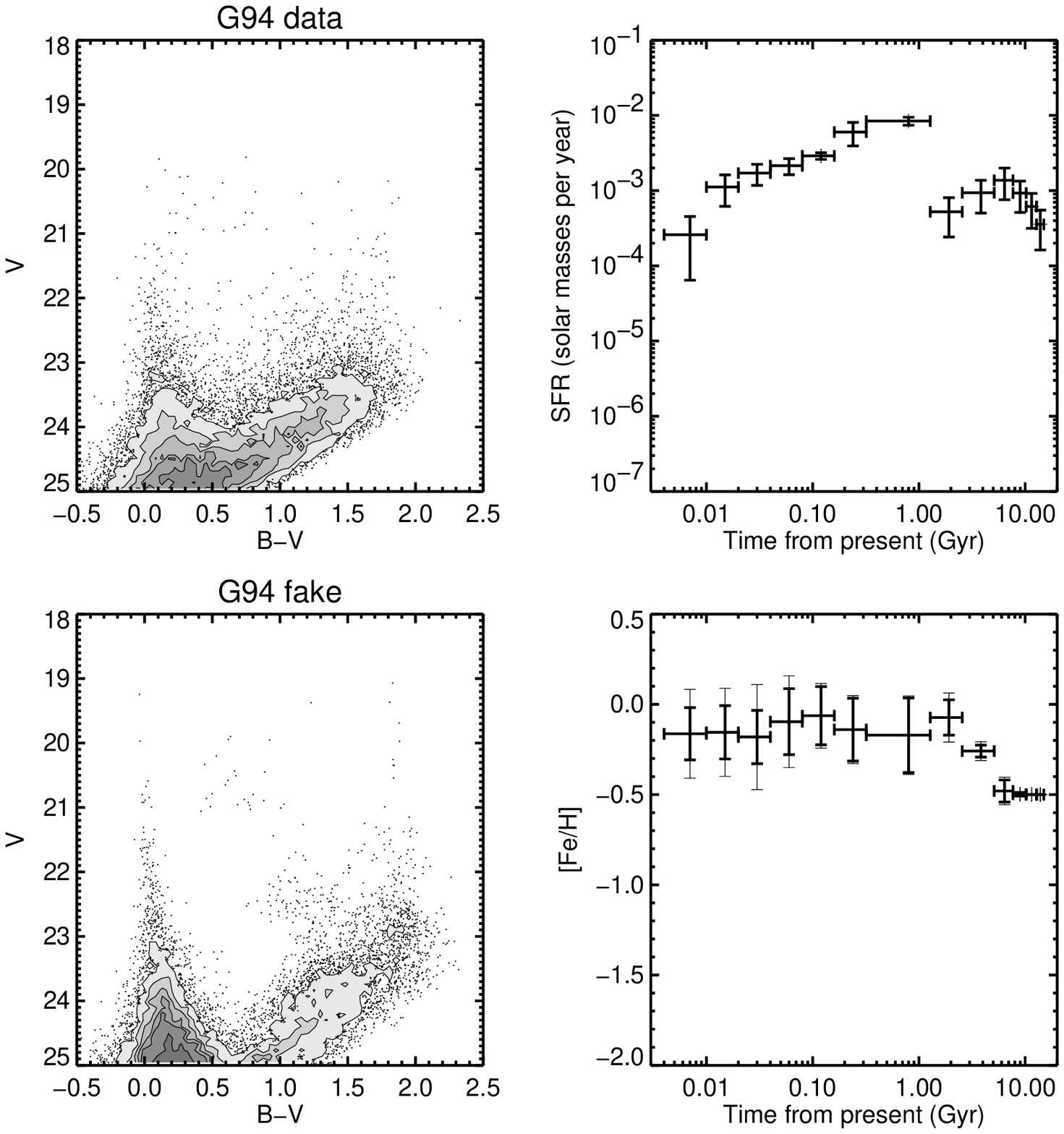,height=6.0in,angle=0}} 
\caption{(k)  The G94 field.}
\end{figure}

\begin{figure}
\figurenum{5}
\centerline{\psfig{file=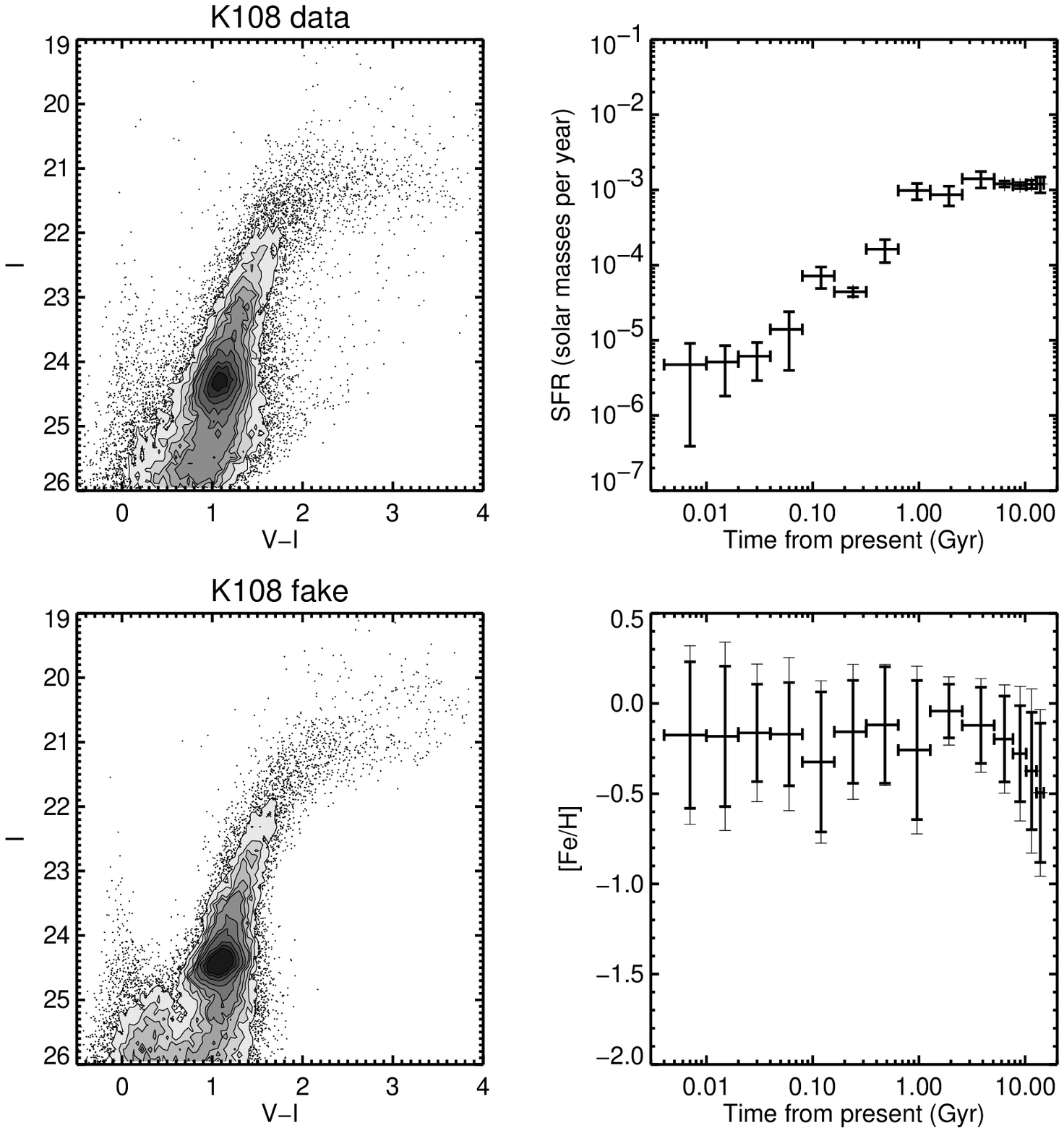,height=6.0in,angle=0}} 
\caption{(l)  The K108 field.}
\end{figure} 

\begin{figure}
\figurenum{5}
\centerline{\psfig{file=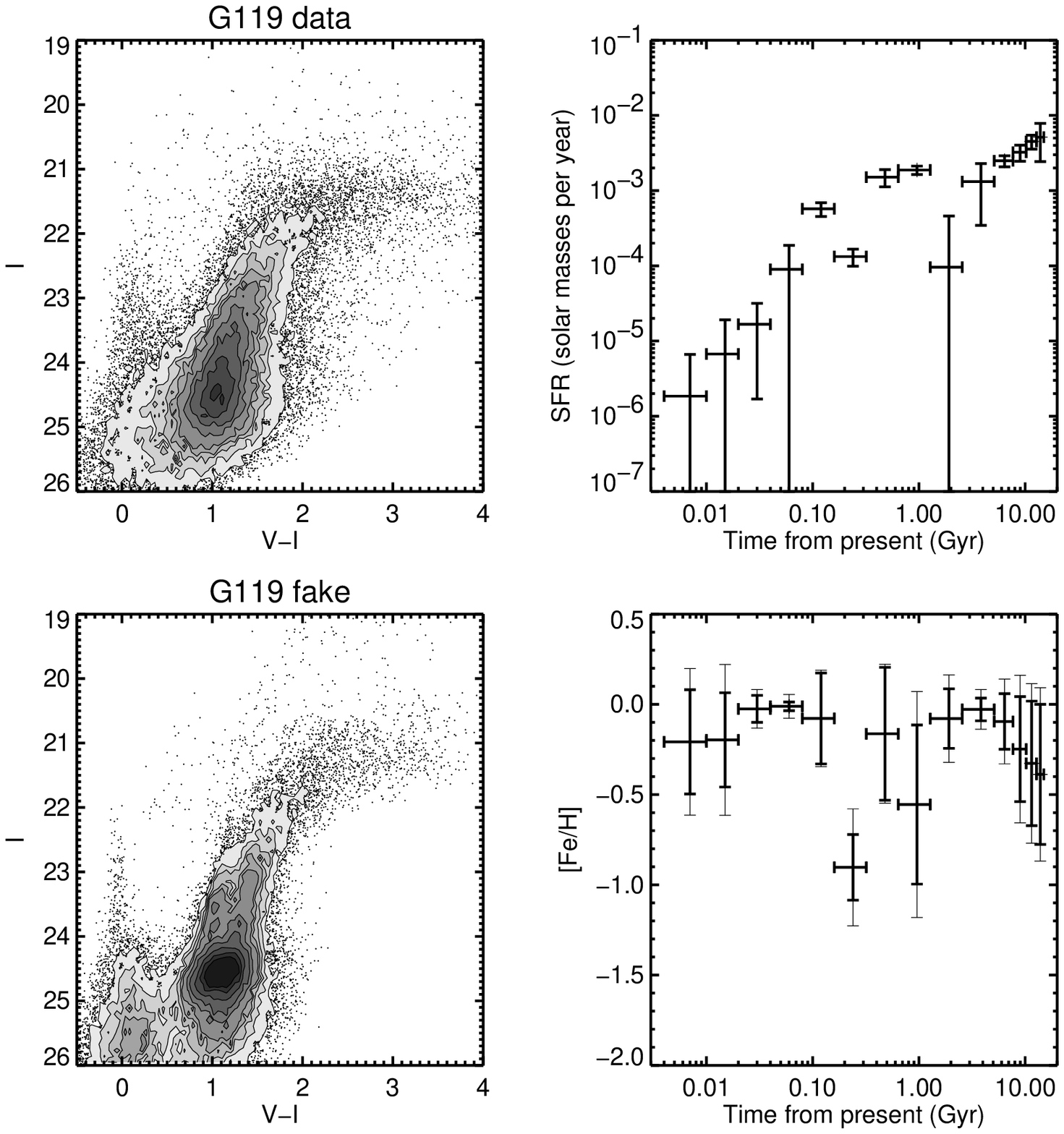,height=6.0in,angle=0}} 
\caption{(m)  The G119 field.}
\end{figure}

\begin{figure}
\figurenum{5}
\centerline{\psfig{file=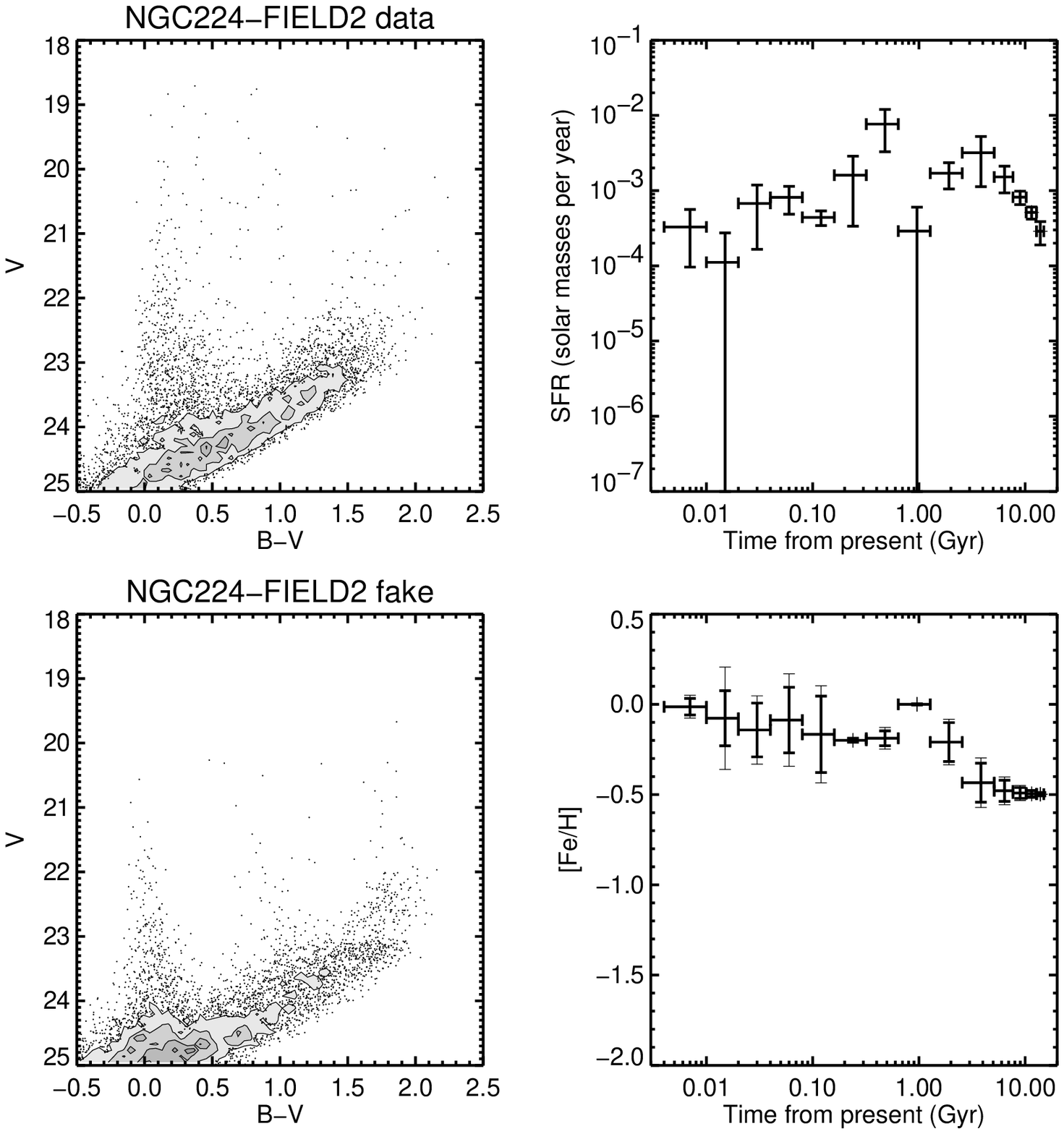,height=6.0in,angle=0}} 
\caption{(n)  The NGC224-FIELD2 field, which was observed in B, V and I.}
\end{figure}      

\begin{figure}
\figurenum{5}
\centerline{\psfig{file=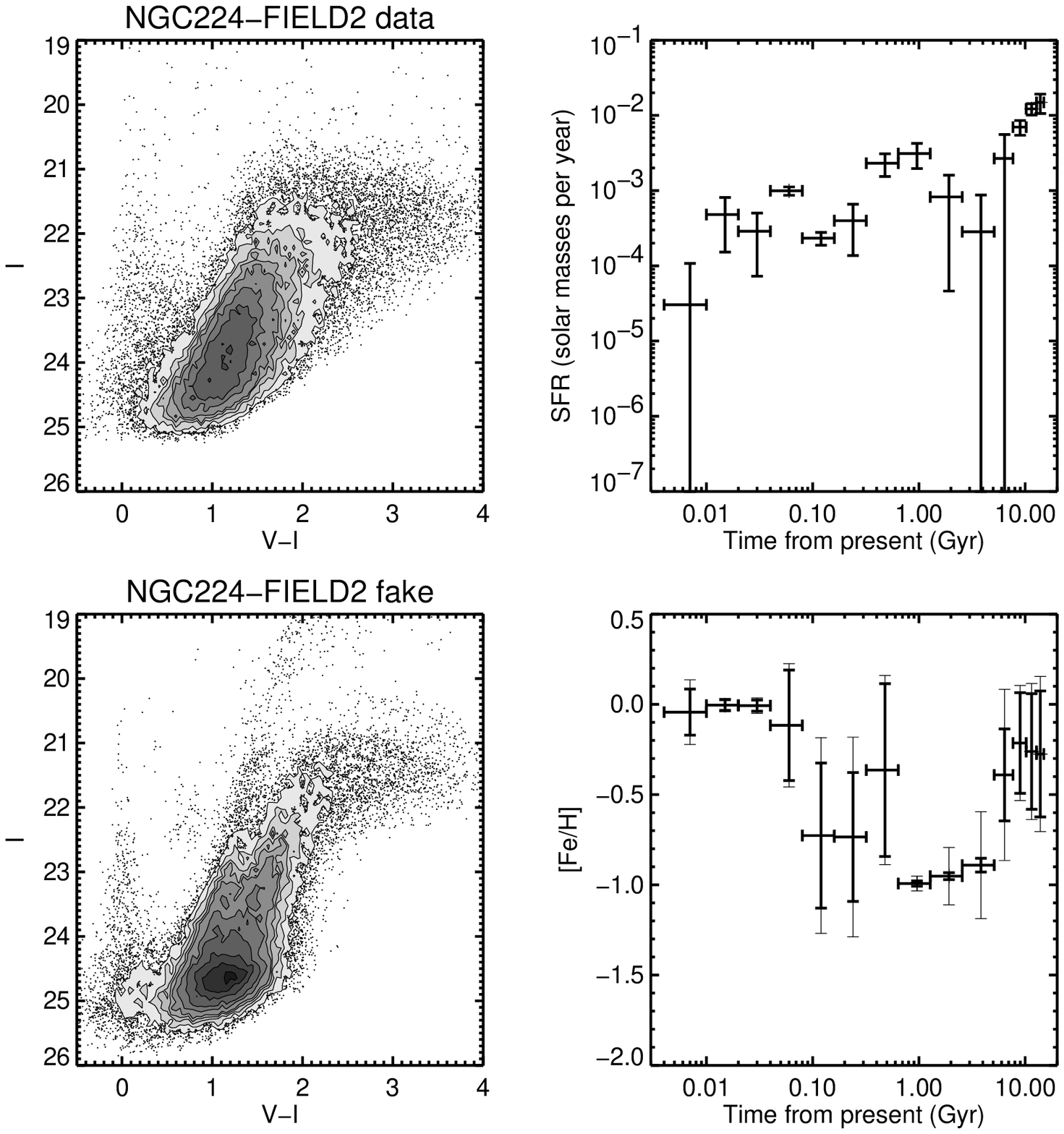,height=6.0in,angle=0}} 
\caption{(o)  The NGC224-FIELD2 field, which was observed in B, V and I.}
\end{figure}  

\begin{figure}
\figurenum{5}
\centerline{\psfig{file=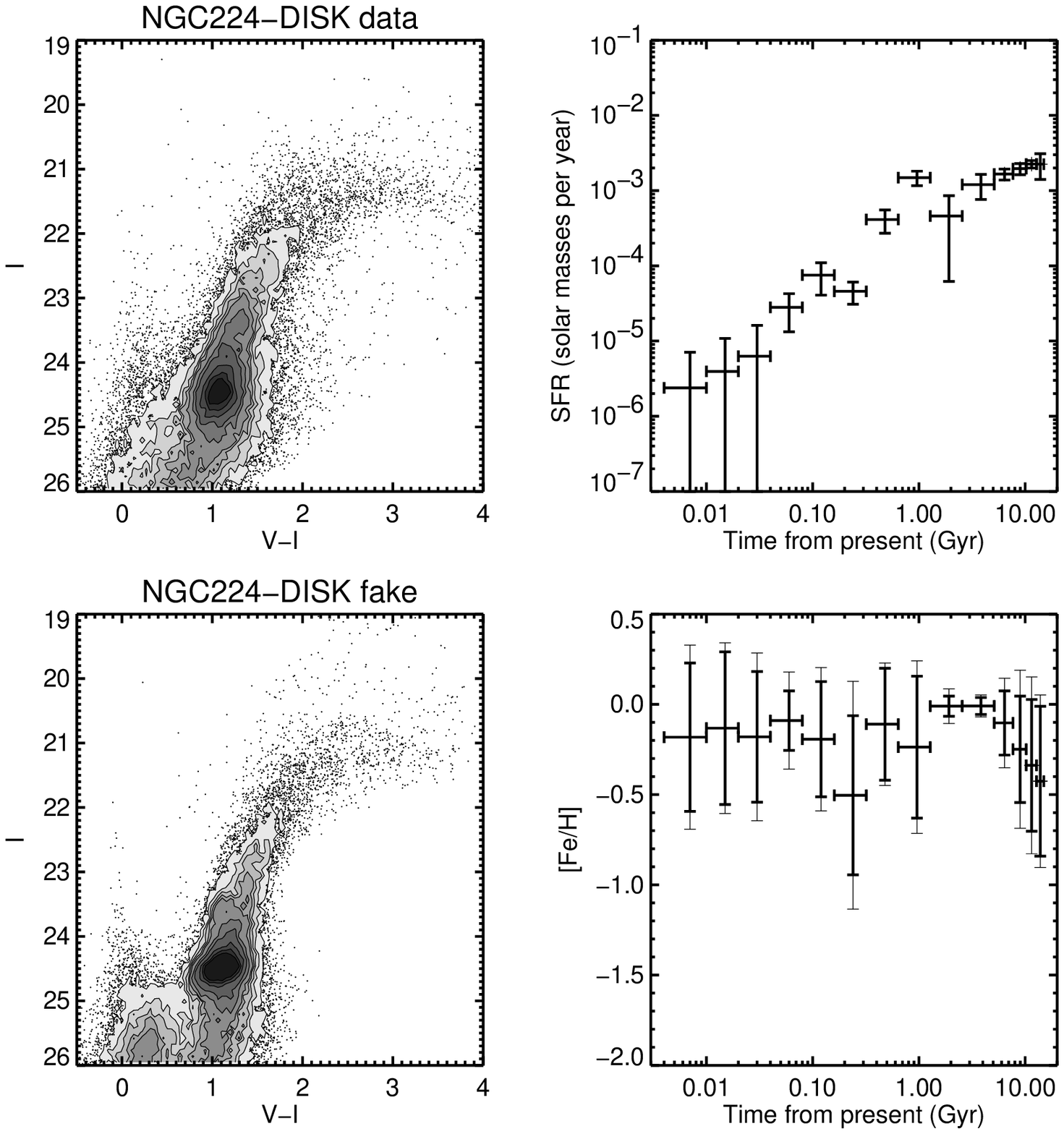,height=6.0in,angle=0}} 
\caption{(p)  The NGC224-DISK field.}
\end{figure} 

\begin{figure}
\figurenum{5}
\centerline{\psfig{file=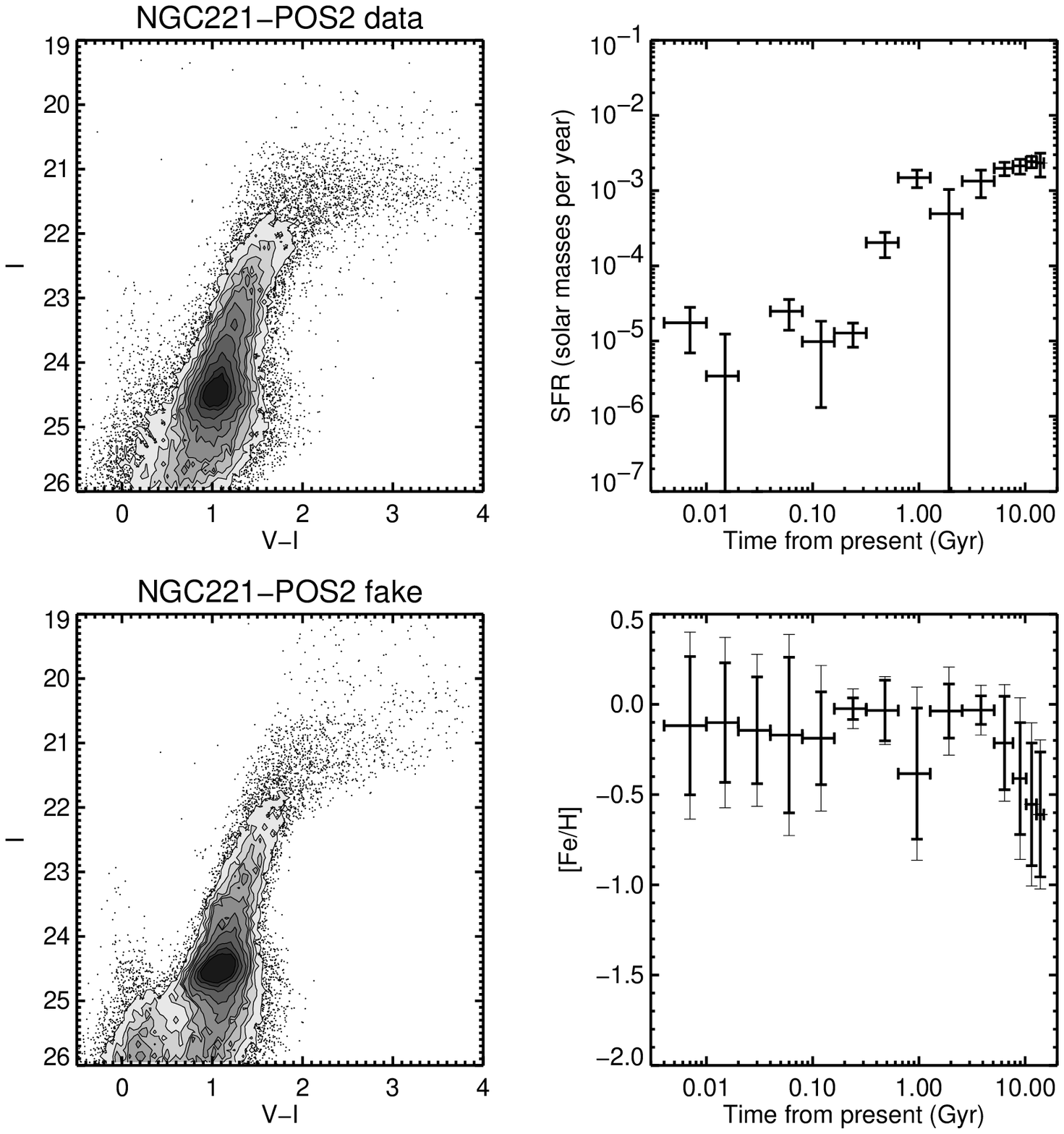,height=6.0in,angle=0}} 
\caption{(q)  The NGC221-POS2 field.}
\end{figure}     

\begin{figure}
\figurenum{5}
\centerline{\psfig{file=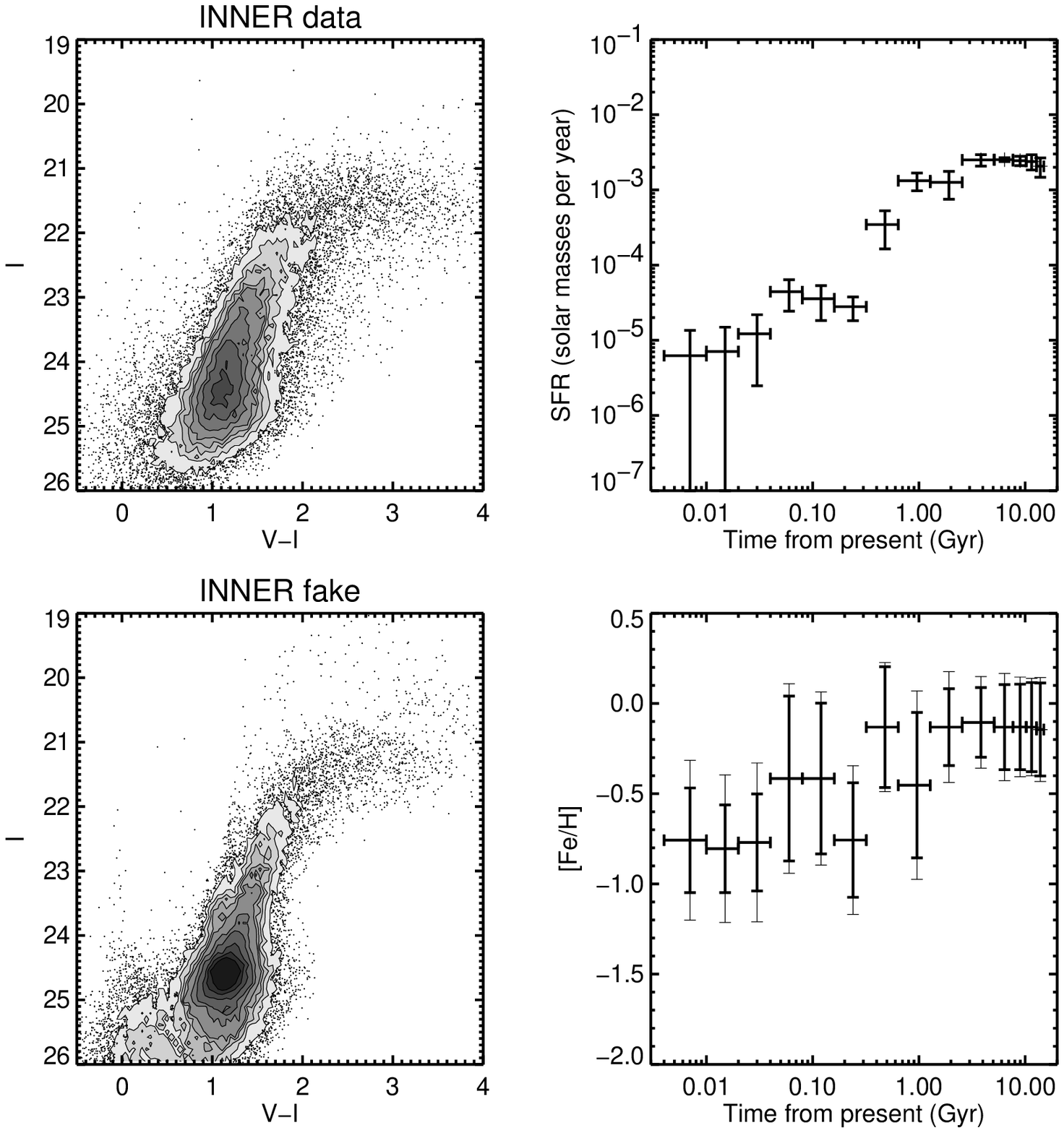,height=6.0in,angle=0}} 
\caption{(r)  The INNER field.}
\end{figure}   

\begin{figure}
\figurenum{5}
\centerline{\psfig{file=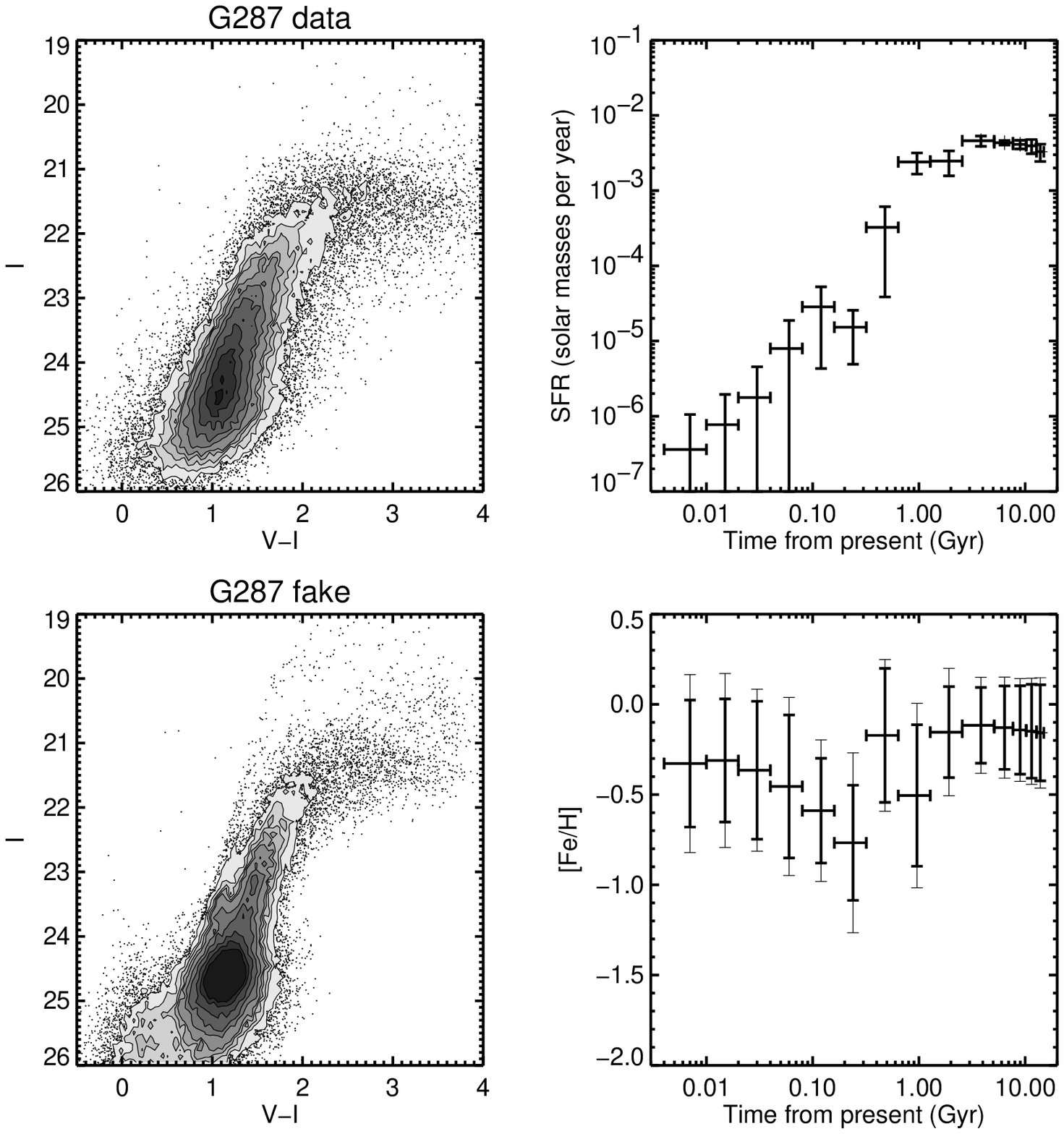,height=6.0in,angle=0}} 
\caption{(s)  The G287 field.}
\end{figure}  

\begin{figure}
\figurenum{5}
\centerline{\psfig{file=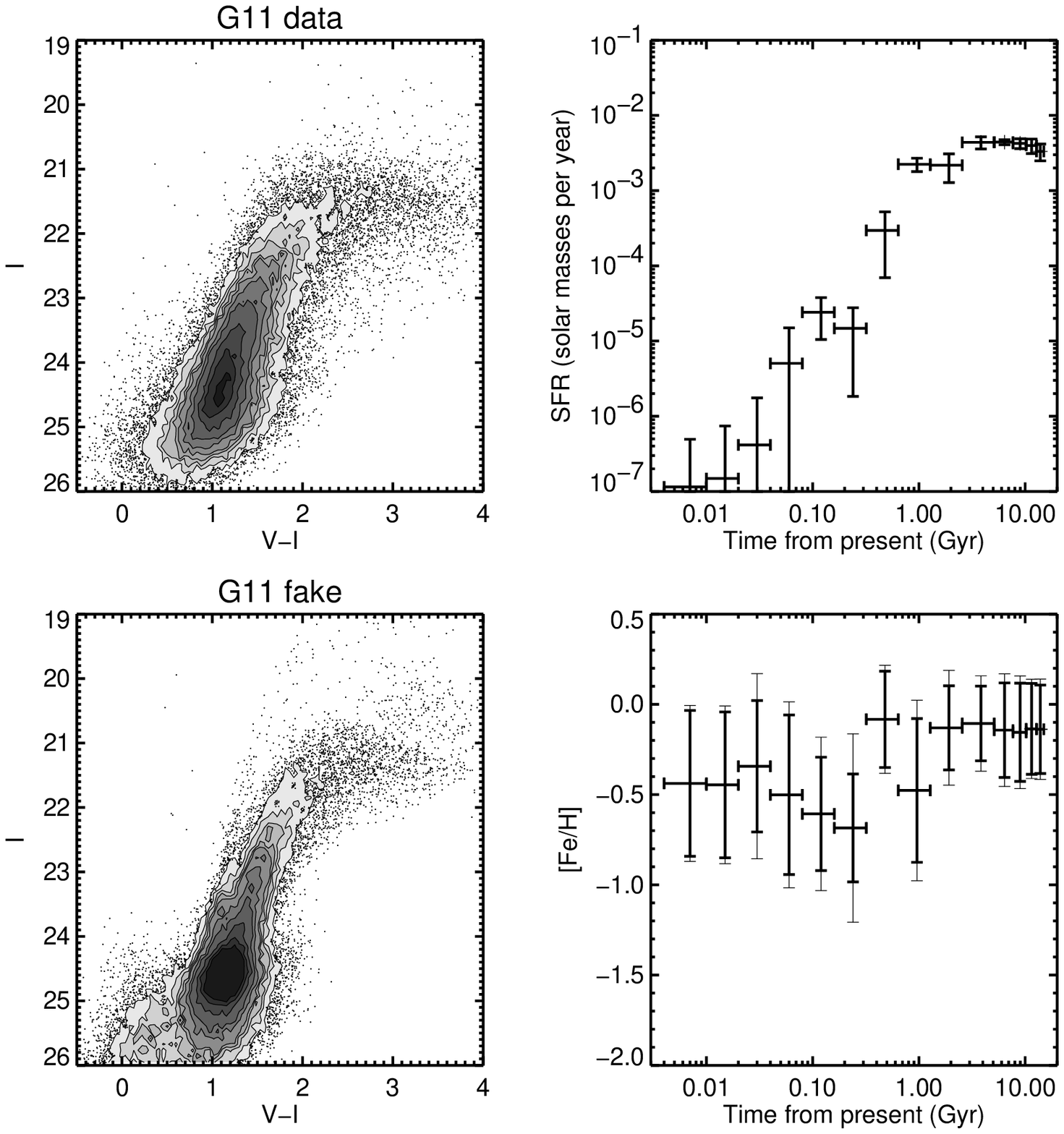,height=6.0in,angle=0}} 
\caption{(t)  The G11 field.}
\end{figure}  

\begin{figure}
\figurenum{5}
\centerline{\psfig{file=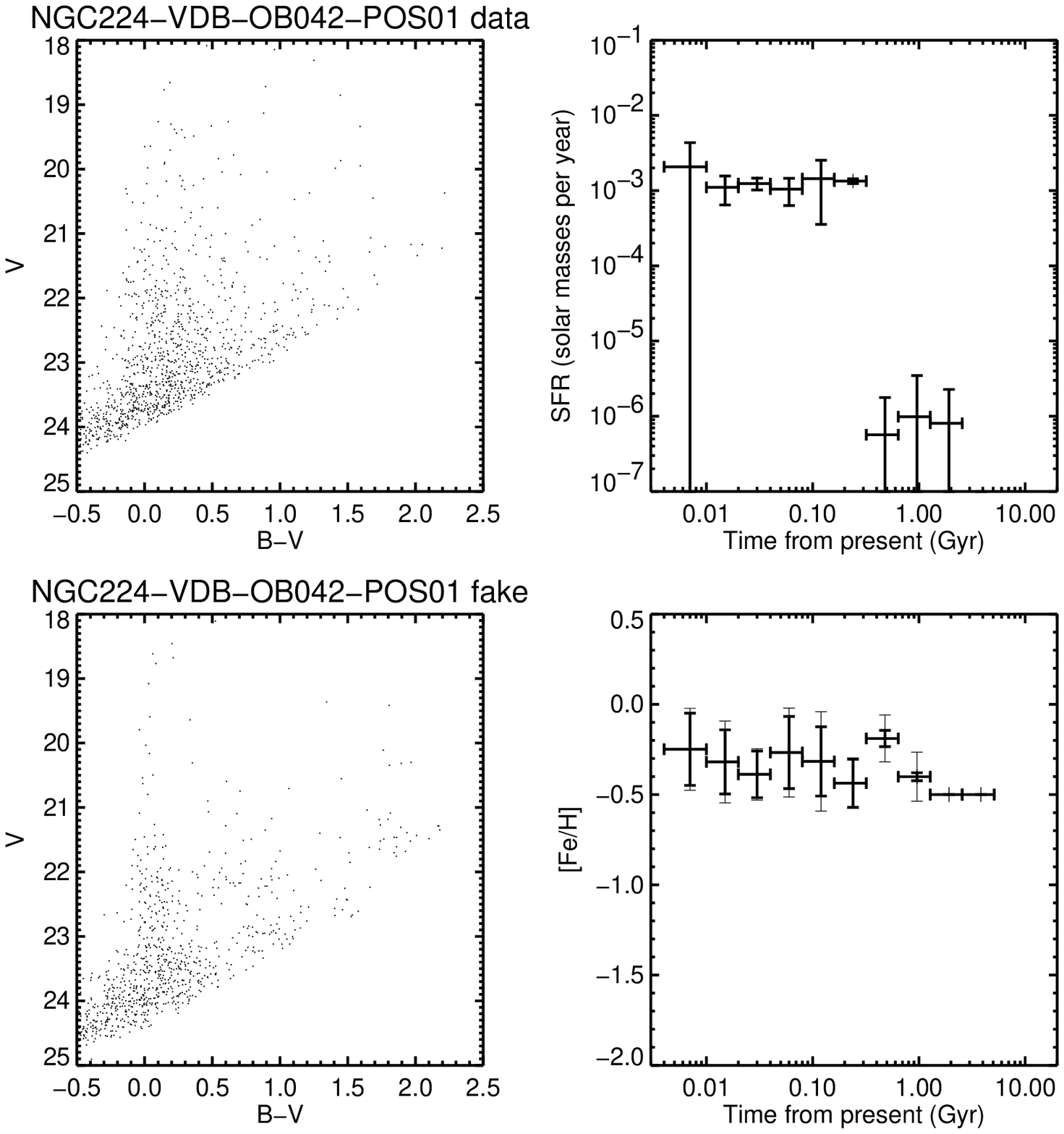,height=6.0in,angle=0}} 
\caption{(u)  The NGC224-VDB-OB042-POS01 field.}
\end{figure}       

\begin{figure}
\figurenum{5}
\centerline{\psfig{file=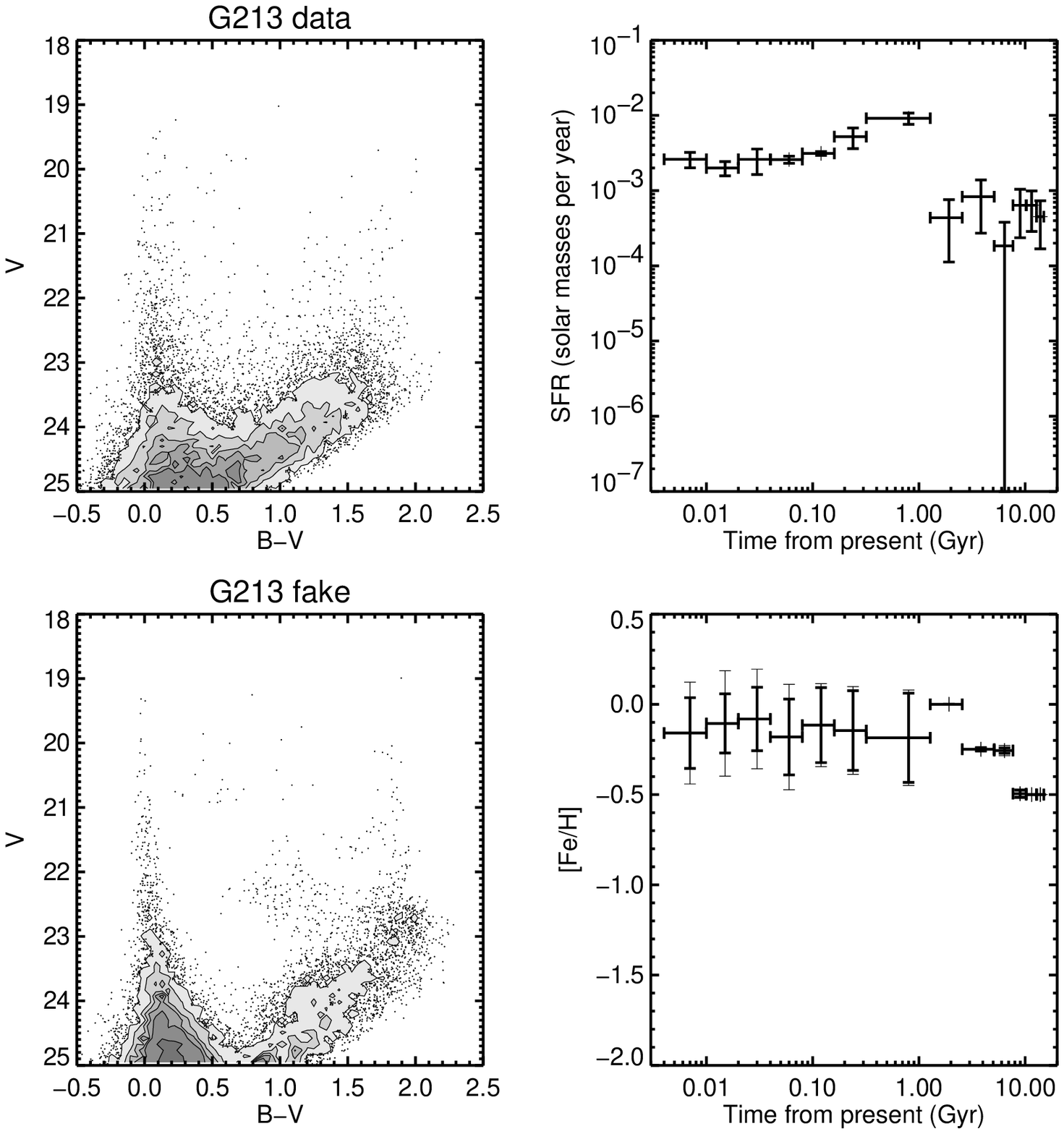,height=6.0in,angle=0}} 
\caption{(v)  The G213 field.}
\end{figure}   

\begin{figure}
\figurenum{5}
\centerline{\psfig{file=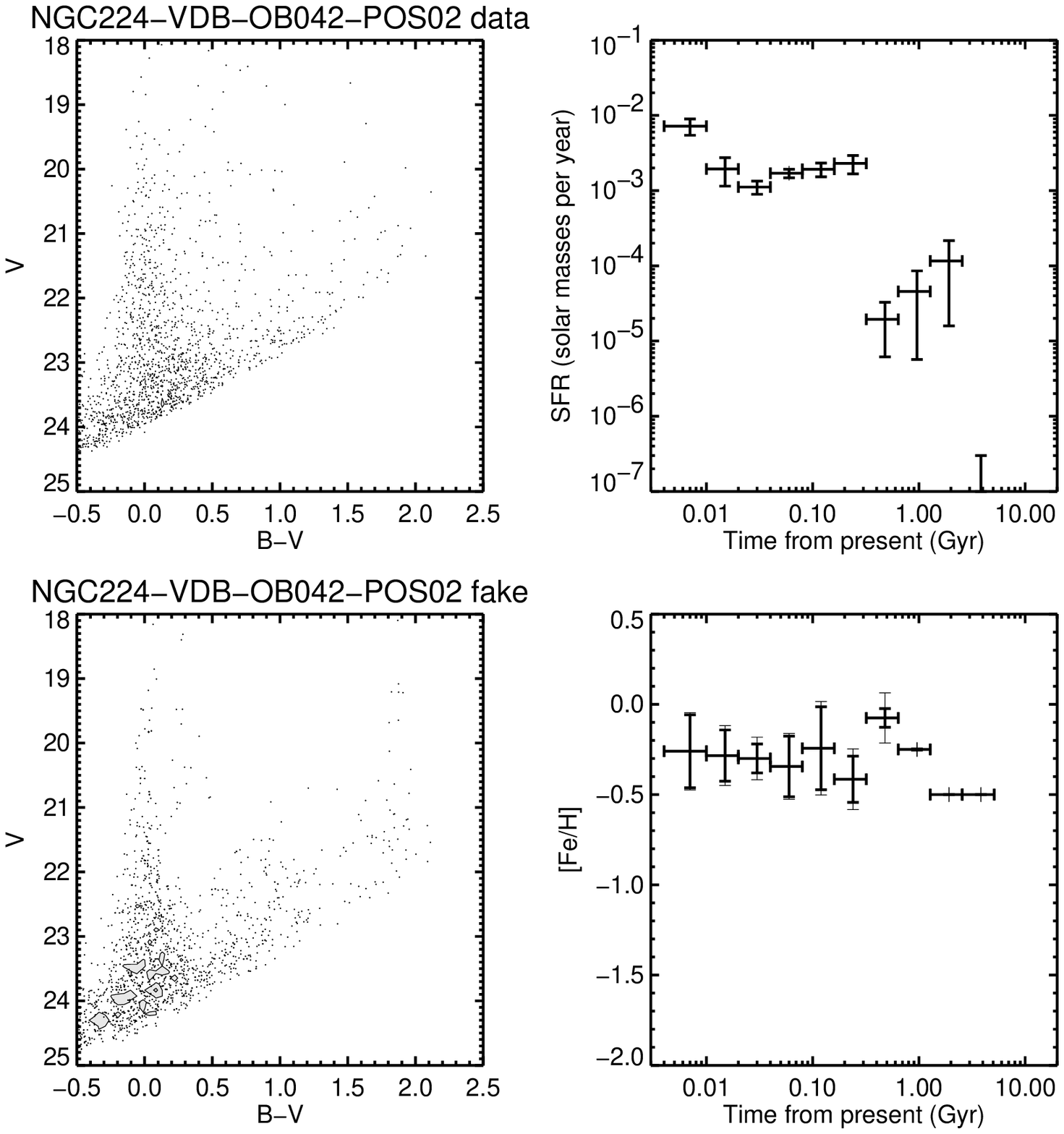,height=6.0in,angle=0}} 
\caption{(w)  The NGC224-VDB-OB042-POS02 field.}
\end{figure}   

\begin{figure}
\figurenum{5}
\centerline{\psfig{file=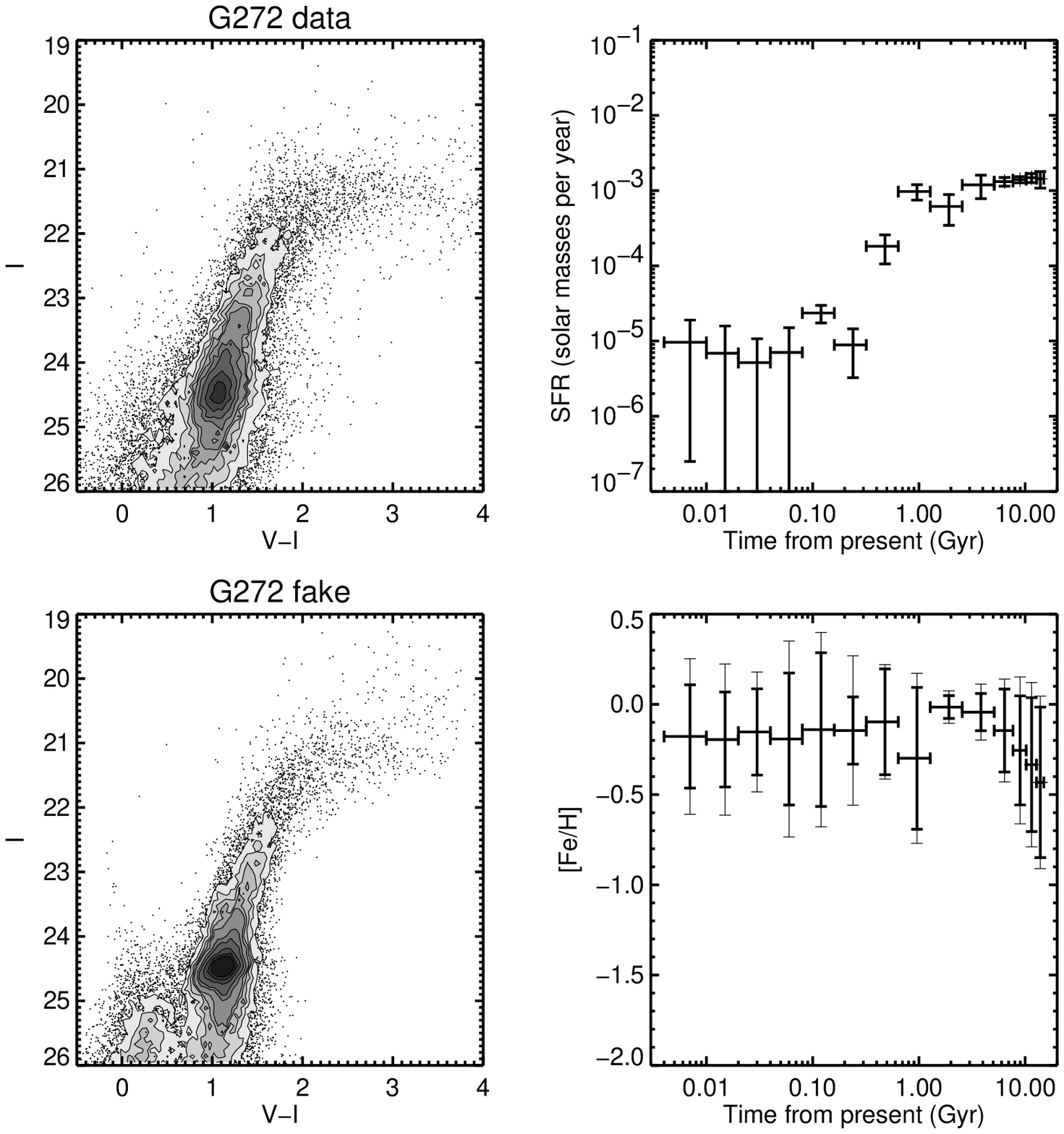,height=6.0in,angle=0}} 
\caption{(x)  The G272 field.}
\end{figure}   

\begin{figure}
\figurenum{5}
\centerline{\psfig{file=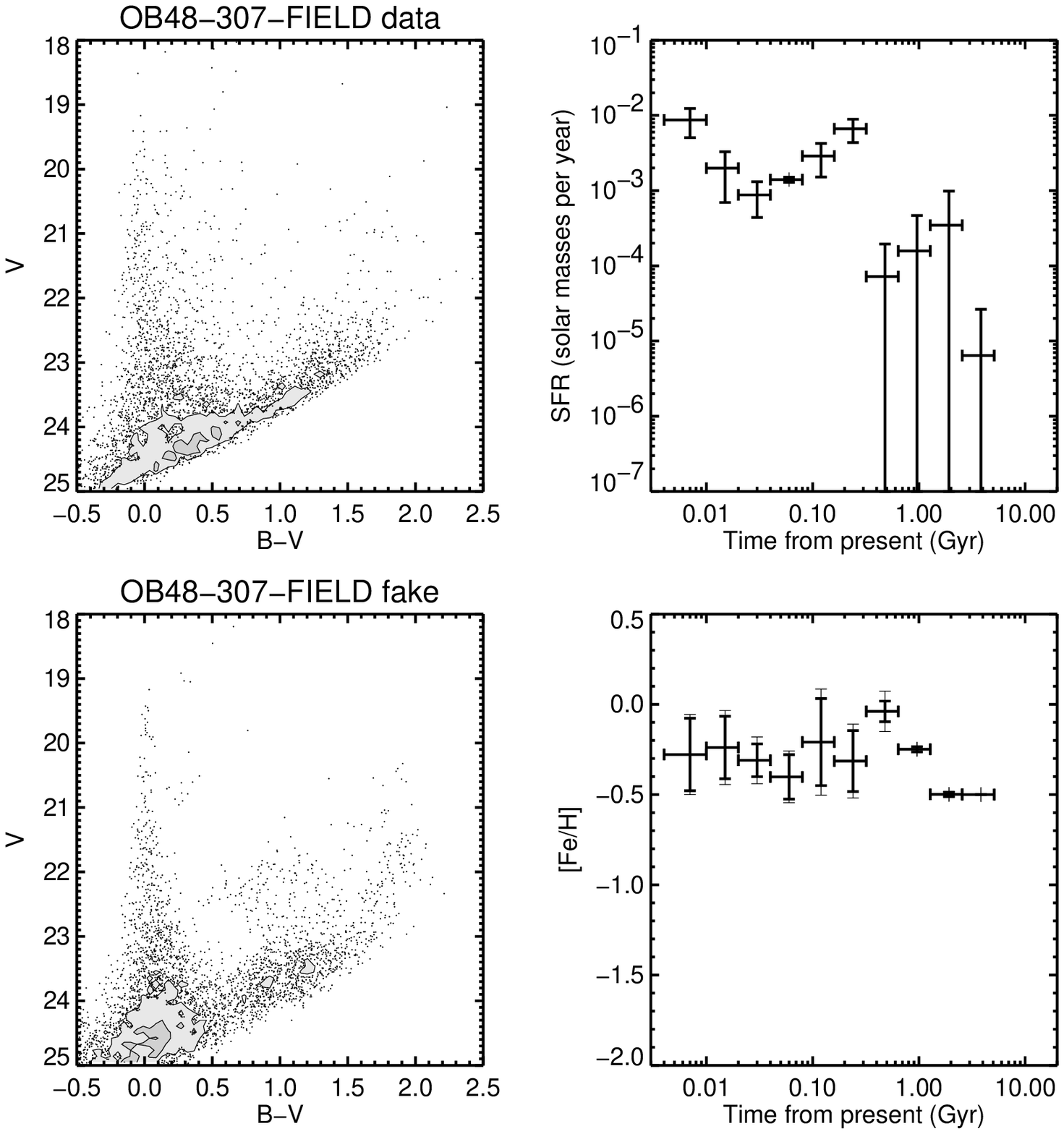,height=6.0in,angle=0}} 
\caption{(y)  The OB48-307-FIELD field.}
\end{figure}   

\clearpage

\begin{figure}
\figurenum{5}
\centerline{\psfig{file=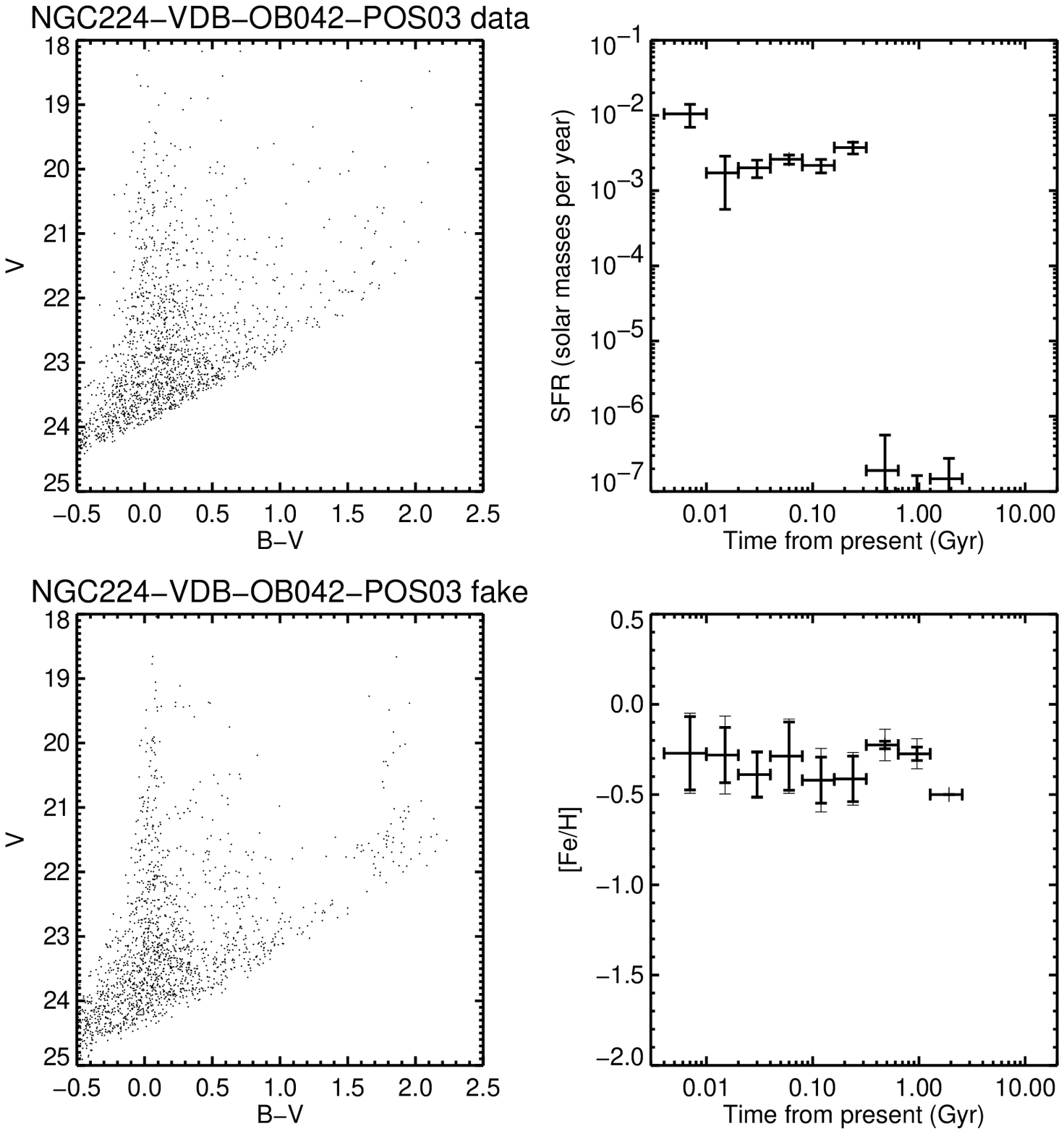,height=6.0in,angle=0}} 
\caption{(z)  The NGC224-VDB-OB042-POS03 field.}
\end{figure}   

\begin{figure}
\figurenum{5}
\centerline{\psfig{file=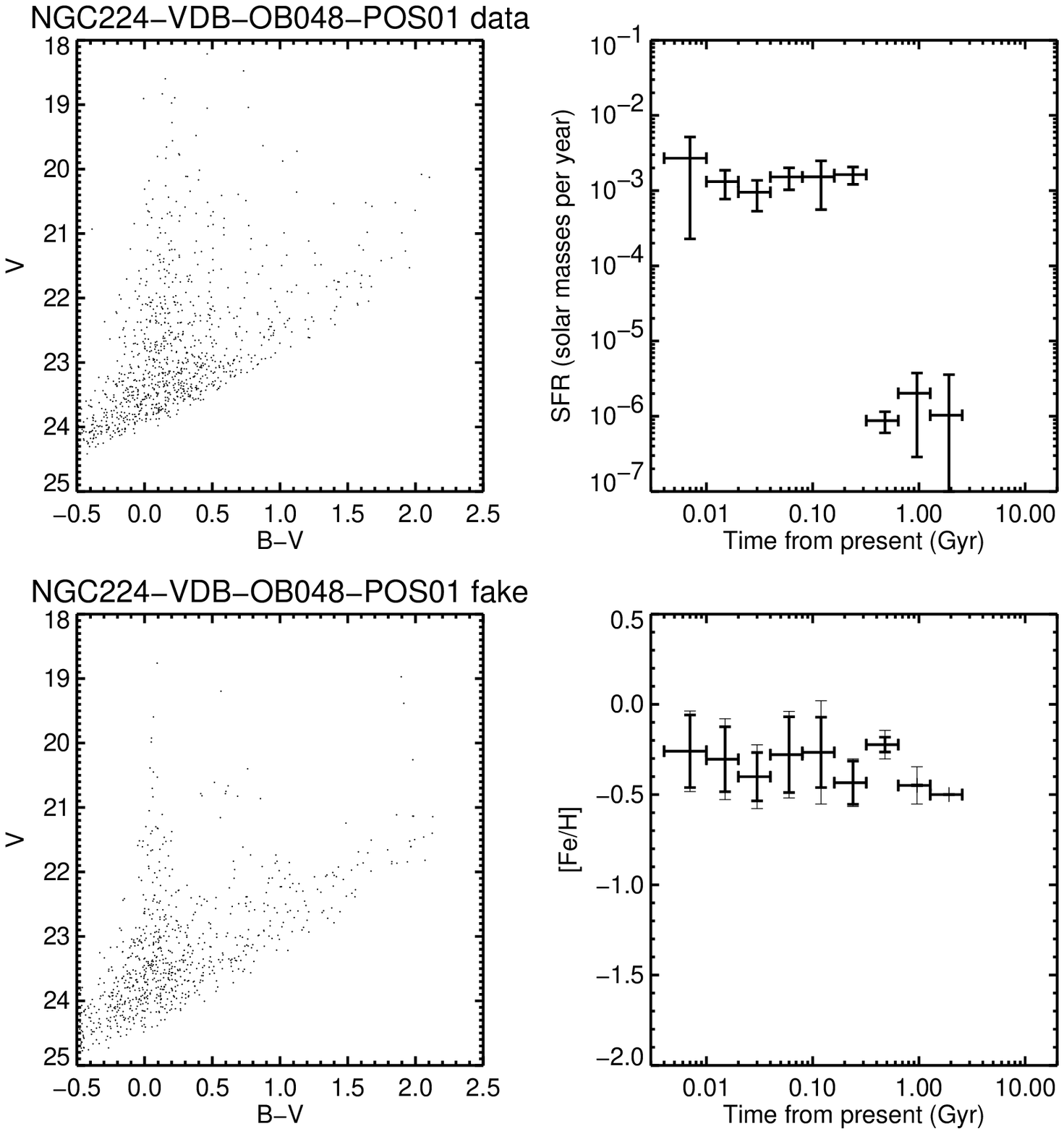,height=6.0in,angle=0}} 
\caption{(aa)  The NGC224-VDB-OB048-POS01 field.}
\end{figure}     

\begin{figure}
\figurenum{5}
\centerline{\psfig{file=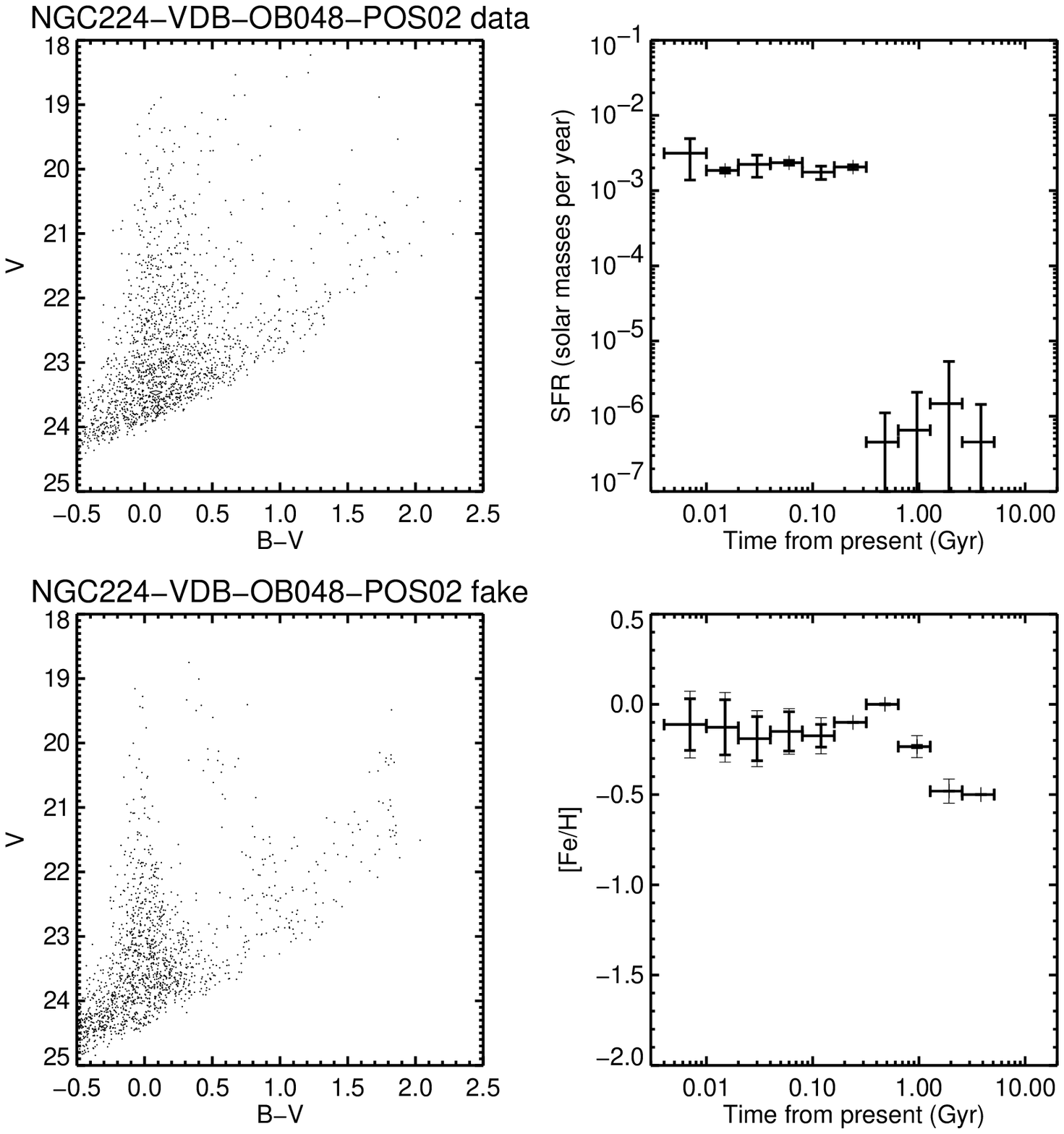,height=6.0in,angle=0}} 
\caption{(ab)  The NGC224-VDB-OB048-POS02 field.}
\end{figure}        

\begin{figure}
\figurenum{5}
\centerline{\psfig{file=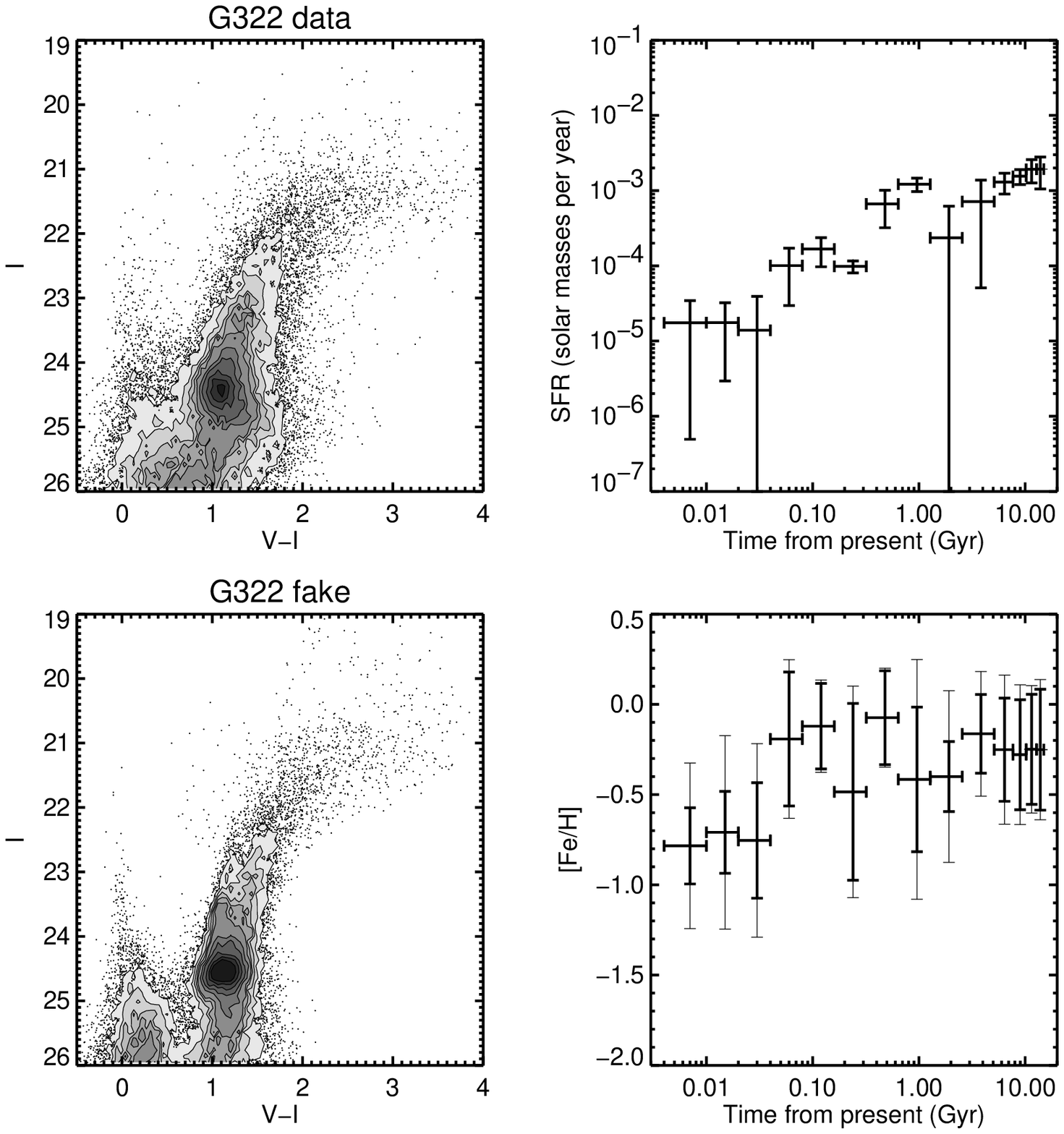,height=6.0in,angle=0}} 
\caption{(ac)  The G322 field.}
\end{figure}        
 
\end{document}